\documentclass[journal]{IEEEtran} 

\usepackage[tbtags]{amsmath}
\usepackage{amsfonts}
\usepackage{amssymb}

\usepackage{amsmath}
\usepackage{graphicx}
\usepackage{subfigure}
\usepackage{floatflt}
\usepackage{fancybox}
\usepackage{epsfig}
\usepackage{cite}
\usepackage{calc}
\usepackage{color}

\newtheorem{thm}{Theorem}[section]
\newtheorem{cor}[thm]{Corollary}

\newtheorem{lem}[thm]{Lemma}
\newtheorem{note}{Remark}
\newtheorem{eg}{Example}
\newtheorem{defn}{Definition}
\newcommand{\bcor}{\begin{cor}}
\newcommand{\ecor}{\end{cor}}
\newcommand{\blem}{\begin{lem}}
\newcommand{\elem}{\end{lem}}
\newcommand{\bthm}{\begin{thm}}
\newcommand{\ethm}{\end{thm}}
\newcommand{\bpf}{\begin{proof}}
\newcommand{\epf}{\end{proof}}
\newcommand{\bit}{\begin{itemize}}
\newcommand{\eit}{\end{itemize}}
\newcommand{\ben}{\begin{enumerate}}
\newcommand{\een}{\end{enumerate}}
\newcommand{\beq}{\begin{equation}}
\newcommand{\eeq}{\end{equation}}
\newcommand{\beqn}{\begin{equation*}}
\newcommand{\eeqn}{\end{equation*}}
\newcommand{\beqa}{\begin{eqnarray}}
\newcommand{\eeqa}{\end{eqnarray}}
\newcommand{\bea}{\begin{eqnarray}}
\newcommand{\eea}{\end{eqnarray}}
\newcommand{\beqan}{\begin{eqnarray*}}
\newcommand{\eeqan}{\end{eqnarray*}}
\newcommand{\bean}{\begin{eqnarray*}}
\newcommand{\eean}{\end{eqnarray*}}

\title{Low Correlation Sequences over the QAM Constellation
}

\author{M. Anand, {\em Student Member, IEEE}, and P. Vijay Kumar, {\em Fellow, IEEE}
\thanks{M. Anand is with the Department of Electrical and Computer Engineering, and
the Coordinated Science Laboratory, University of Illinois, Urbana,
IL 61801 USA (email:amurali2@uiuc.edu).  This work was carried out
while M. Anand was with the Department of Electrical Communication
Engineering, Indian Institute of Science Bangalore, 560 012 India.}
\thanks{P. Vijay Kumar is with the Department of EE-Systems, University of
Southern California, Los Angeles, CA 90089 USA (email:
vijayk@usc.edu).  This work was carried out while P. Vijay Kumar was
on leave of absence at the Indian Institute of Science, Bangalore.}
\thanks{This research is supported in part by NSF-ITR CCR-0326628 and in part by
the DRDO-IISc Program on Advanced Research in Mathematical
Engineering.} }

\begin{document}
\maketitle\thispagestyle{empty}
\bibliographystyle{ieeetran}

\begin{abstract}

This paper presents the first concerted look at low correlation
sequence families over QAM constellations of size $M^2=4^m$ and
their potential applicability
as spreading sequences
in a CDMA setting.
 Five constructions
are presented, and it is shown how such sequence families have the
ability to transport a larger amount of data as well as enable
variable-rate signalling on the reverse link.

Canonical family ${\cal CQ}$ has period $N$, normalized
maximum-correlation parameter $\overline{\theta}_{\max}$ bounded
above by $ \lesssim a \, \sqrt{N}$, where $a$ ranges from $1.8$ in
the $16$-QAM case to $3.0$ for large $M$.  In a CDMA setting, each
user is enabled to transfer $2m$ bits of data per period of the
spreading sequence which can be increased to $3m$ bits of data by
halving the size of the sequence family.  The technique used to
construct ${\cal CQ}$ is easily extended to produce larger
sequence families and an example is provided.

Selected family ${\cal SQ}$ has a lower value of $\theta_{\max}$ but
permits only $(m+1)$-bit data modulation.   The interleaved $16$-QAM
sequence family ${\cal IQ}$ has $\overline{\theta}_{\max} \lesssim
\sqrt{2} \, \sqrt{N}$ and supports $3$-bit data modulation.

The remaining two families are over a quadrature-PAM (Q-PAM)
subset of size $2M$ of the $M^2$-QAM constellation.  Family ${\cal
P}$ has a lower value of $\overline{\theta}_{\max}$ in comparison
with Family ${\cal SQ}$, while still permitting $(m+1)$-bit data
modulation.  Interleaved family ${\cal IP}$, over the $8$-ary
Q-PAM constellation, permits $3$-bit data modulation and
interestingly, achieves the Welch lower bound on
$\overline{\theta}_{\max}$.

\end{abstract}

\begin{keywords}
QAM, Q-PAM, low-correlation sequences, CDMA, variable-rate
signalling, quaternary sequences, galois rings.
\end{keywords}

\section{Introduction} \label{sec:intro}

In Direct-Sequence Code Division Multiple Access (DS-CDMA) systems,
low-correlation spreading sequences are employed to separate the
signals of different users. In this paper, constructions of families
of low-correlation spreading sequences over the $M^2$-QAM
constellation, $M = 2^m$, as well as over a quadrature-PAM
subconstellation of size $2M$ are presented.

The periodic correlation between two complex-valued sequences, $\{
s(j, t) \}$ and $\{ s(k, t) \}$, at time shift $\tau$ is
defined\footnote{For the sake of brevity, we abbreviate and use
$s(j),s(k)$ throughout in place of $s(j,t), s(k,t)$ whenever these
terms appear in the subscript.} as \beqan
    \theta_{s(j), \ s(k)}(\tau) & = & \sum_{t = 0}^{N - 1} \,
    s(j, t + \tau) \, \overline{s(k, t)} \\
    & & \hspace{0.3in} \mbox{where} \ 0 \leq \tau \leq (N - 1)
\eeqan with $(t + \tau)$ computed modulo $N$.  This form of
correlation is also referred to as even-periodic correlation to
differentiate it from other forms of correlation between two
sequences.

We define the maximum correlation parameter for a family of
sequences to be \beq
    \theta_{\max} := \max \left\{
    \left| \theta_{s(j), \ s(k)}(\tau) \right|
    \left| \begin{array}{l}
    \mbox{either $j \neq k$} \\
    \mbox{or} \ \tau \neq 0
    \end{array} \right.\right\}. \label{eqref:theta_max_old}
\eeq  The parameters commonly used to compare sequence families
are the size of the symbol alphabet, the period $N$ of each
sequence, the number of cyclically-distinct sequences in the
family and the value of $\theta_{\max}$.

\subsection{Motivation} There are several reasons for being
 interested in low-correlation sequences over the QAM and Q-PAM
constellations
\begin{itemize} \item the increasing popularity of the QAM alphabet
for signalling purposes  \item the potential for modulating data at
higher data rates \item the potential for variable-rate signalling
on the reverse link of a CDMA system \item the potential for larger
Euclidean distance between the signals corresponding to different
data bits of the same user, thus improving reliability of
communication \item the larger symbol alphabet that makes the
sequences harder to predict from an intercepted fragment \item the
potential for data modulation in ways that are not transparent to a
casual observer which makes it harder for the casual observer to
recover the data from an observed fragment.
\end{itemize}

When considering high-order modulation, the first approach that
suggests itself is one of ``multiplying'' the QPSK spreading
sequence Family ${\cal A}$ sequence by symbols from the $M^2$-QAM
constellation. This would, however, mean that the transmitted
energy per period of the spreading sequence would vary vastly
depending upon the particular QAM symbol being transmitted. As a
result, each user would experience a varying amount of
interference depending upon the particular combination of symbols
being transmitted by the other users. The QAM symbols with low
magnitude would be far more susceptible to interference than those
with larger magnitude. In contrast, in all of the designs
presented here, every spreading sequence has the same energy.

A second alternative one might consider would be to use Family
${\cal A}$ and modulate the code sequence with $M^2$-ary phase
modulation.  This would however, lead to smaller Euclidean
distance between distinct data symbols of the same user and also
make the system more sensitive to phase offsets.

\begin{table*}[!ht]
    \caption{Parameters of Some Relevant Prior Constructions in the Literature} \label{table:others}
\begin{center}
\begin{tabular}{|c|c|c|c|c|c|}
    \hline \hline

    Family & Constellation & Period & Family Size  & Data Rate    &  Asymptotic Upper \\
                                         &  &   &  &  & Bound on $\overline{\theta}_{max}$ \\
    \hline
    $\cal A$ \cite{Sol,BozThesis,BozHamKum}& $4$-PSK    & $N = 2^r - 1$ & $N + 2$ & 2 &
    $\sqrt{N}$\\
    \hline
    ${\cal{S}}(1)$ \cite{KumHelCalHam}& $4$-PSK    & $N = 2^r - 1$ & $\geq (N + 1)^2$
    & 2 & $2 \, \sqrt{N}$\\
    \hline
    ${\cal{S}}(2)$ \cite{KumHelCalHam}& $4$-PSK    & $N = 2^r - 1$ & $ \geq (N + 1)^3$
    & 2 & $4 \, \sqrt{N}$\\
    \hline
    ${\cal{S}}(p)$ \cite{KumHelCalHam}& $4$-PSK    & $N = 2^r - 1$
    & $\geq (N + 1)^{p + 1}$ & 2 & $2^p \, \sqrt{N}$\\
    \hline \hline
       & $4$-PSK    & $N = 2^r - 1$ & $\geq (N + 1)^2$ & 2 &
    $2 \, \sqrt{N}$\\
    \cline{2-6}
    WCU Sequences \cite{KumHelCal} & $8$-PSK    & $N = 2^r - 1$ & $\geq (N + 1)^3$ & 3 &
    $3 \, \sqrt{N}$\\
    \cline{2-6}
        & $8$-PSK   & $N = 2^r - 1$ & $\geq (N + 1)^4$ & 3 &
    $4 \, \sqrt{N}$\\
    \cline{2-6}
        & $8$-PSK   & $N = 2^r - 1$ & $\geq (N + 1)^5$ & 3 &
    $5 \, \sqrt{N}$\\
    \cline{2-6}
        & $8$-PSK   & $N = 2^r - 1$ & $\geq (N + 1)^6$ & 3 &
    $6 \, \sqrt{N}$\\
    \hline
    ${\cal Q}_{B}$ \cite{Boz} & $16$-QAM & $N = 2^r - 1$ & $(N + 1)/2$ &(not  &
    $1.8 \sqrt{N}$ \\
    & & & & discussed) & (derived here) \\
    \hline \hline
\end{tabular}
\end{center}
\end{table*}

\begin{table*}[!ht]
    \caption{Parameters of the Family ${\cal CQ}_{M^2}$ for various $M$}
    \label{table:seq_qam_nom}
\begin{center}

\begin{tabular}{|c|c|c|c|c|c|}
    \hline \hline

    Family & $M$ & Constellation &  Family Size  & Data Rate  & Asymptotic Upper \\
              & &  &   &  & Bound on $\overline{\theta}_{max}$ \\
    \hline
    ${\cal CQ}_{16}$ & $4$ & $16$-QAM   & $(N + 1)/2$ & 4   &
    $1.8 \, \sqrt{N}$\\
    \hline
    ${\cal CQ}_{64}$ & $8$ & $64$-QAM    & $\lfloor (N + 1)/3 \rfloor$ & 6   &
    $2.33 \, \sqrt{N}$\\
    \hline
    ${\cal CQ}_{256}$ & $16$ & $256$-QAM   & $(N + 1)/4$ & 8   &
    $2.41 \, \sqrt{N}$\\
    \hline
    ${\cal CQ}_{M^2}$ & $M$ & $M^2$-QAM   & $\lfloor (N + 1)/m \rfloor$ & $2m$   &
    $3 \, \sqrt{N}$\\
    \hline \hline
\end{tabular}
\end{center}
\end{table*}

\begin{table*}[!ht]
    \caption{Parameters of the Family ${\cal CQ}_{M^2}$ with increased data rate}
    \label{table:seq_qam_nom_data}
\begin{center}

\begin{tabular}{|c|c|c|c|c|c|}
    \hline \hline

    Family & $M$ & Constellation &  Family Size  & Data Rate  & Asymptotic Upper \\
              & &  &   &  & Bound on $\overline{\theta}_{max}$ \\
    \hline
    ${\cal CQ}_{16}$ & $4$ & $16$-QAM   & $(N + 1)/4$ & $6$   &
    $1.8 \, \sqrt{N}$\\
    \hline
    ${\cal CQ}_{64}$ & $8$ & $64$-QAM    & $\lfloor (N + 1)/6 \rfloor$ & $9$   &
    $2.33 \, \sqrt{N}$\\
    \hline
    ${\cal CQ}_{256}$ & $16$ & $256$-QAM   & $(N + 1)/8$ & $12$   &
    $2.41 \, \sqrt{N}$\\
    \hline
    ${\cal CQ}_{M^2}$ & $M$ & $M^2$-QAM   & $\lfloor (N + 1)/2m \rfloor$ & $3m$   &
    $3 \, \sqrt{N}$\\
    \hline \hline
\end{tabular}
\end{center}
\end{table*}

\begin{table*}[!ht]
    \caption{Parameters of the Family ${\cal SQ}_{M^2}$ for various $M$}
    \label{table:seq_qam}
\begin{center}

\begin{tabular}{|c|c|c|c|c|c|}
    \hline \hline

    Family & $M$ & Constellation &  Family Size  & Data Rate & Asymptotic Upper \\
              & &  &   &  & Bound on $\overline{\theta}_{max}$ \\
    \hline
    ${\cal SQ}_{16}$ & $4$ & $16$-QAM    & $(N + 1)/2$ & 3   &
    $1.61 \, \sqrt{N}$\\
    \hline
    ${\cal SQ}_{64}$ & $8$ & $64$-QAM    & $(N + 1)/4$ & 4   &
    $2.10 \, \sqrt{N}$\\
    \hline
    ${\cal SQ}_{256}$ & $16$ & $256$-QAM   & $\geq (N + 1)/8$ & 5   &
    $2.41 \, \sqrt{N}$\\
    \hline
    ${\cal SQ}_{1024}$ & $32$ & $1024$-QAM   & $\geq (N + 1)/8$ & 6   &
    $2.58 \, \sqrt{N}$\\
    \hline
    ${\cal SQ}_{M^2}$ & $M$ & $M^2$-QAM   &
    $\geq (N + 1)/4m$ & $m + 1$   &
    $2.76 \, \sqrt{N}$\\
    \hline \hline
\end{tabular}
\end{center}
\end{table*}

\begin{table*}[!ht]
    \caption{Parameters of the Particular $16$-QAM sequence family ${\cal IQ}_{16}$.}
    \label{tab:qam_16}
\begin{center}
\begin{tabular}{|c|c|c|c|c|c|c|}
    \hline \hline
    Family & $M$ & Constellation  & Period  & Family Size  & Data Rate  &  Asymptotic Upper \\
    & & & & & &
    Bound  on $\overline{\theta}_{max}$ \\
    \hline
    ${\cal IQ}_{16}$ & $4$ & $16$-QAM   & $N = 2(2^r - 1)$ & $(N + 2)/4$ & 3 &
    $\sqrt{2} \sqrt{N}$ \\
    \hline \hline
\end{tabular}
\end{center}
\end{table*}

\begin{table*}[!ht]
     \caption{Parameters of the Family ${\cal P}_{2M}$ for various $M$}
    \label{table:seq_am_psk}
\begin{center}

\begin{tabular}{|c|c|c|c|c|c|}
    \hline \hline

    Family & $M$ & Constellation &  Family Size  & Data Rate  & Asymptotic Upper \\
              &  &  &   &  & Bound on $\overline{\theta}_{max}$ \\
    \hline
    ${\cal P}_{8}$ & $4$ & $8$-ary Q-PAM    & $(N + 1)/2$ & 3  &
    $1.34 \, \sqrt{N}$\\
    \hline
    ${\cal P}_{16}$ & $8$ & $16$-ary Q-PAM    & $(N + 1)/4$ & 4  &
    $1.72 \, \sqrt{N}$\\
    \hline
    ${\cal P}_{32}$ & $16$ & $32$-ary Q-PAM   & $\geq (N + 1)/7$ and $\leq (N+1)/5$& 5   &
    $1.96 \, \sqrt{N}$\\
    \hline
    ${\cal P}_{64}$ & $32$ & $64$-ary Q-PAM   & $\geq (N + 1)/10$ and $\leq (N+1)/6$ & 6   &
    $2.09 \, \sqrt{N}$\\
    \hline
    ${\cal P}_{2M}$ & $M$ & $2M$-ary Q-PAM   & $\geq (N + 1)/(m^2)$ & $m + 1$   &
    $\sqrt{5} \, \sqrt{N}$\\
    \hline \hline
\end{tabular}
\end{center}
\end{table*}

\begin{table*}[!ht]
    \caption{Parameters of the Welch-Bound-Achieving $8$-ary Q-PAM sequence family ${\cal IP}_{8}$.}
    \label{table:opt}
\begin{center}
\begin{tabular}{|c|c|c|c|c|c|c|c|c|c|}
    \hline \hline
    Family & $M$ & Constellation  & Period  & Family Size  &  Data Rate  &  Asymptotic Upper \\
     & & & & & & on $\overline{\theta}_{max}$ \\
    \hline
    \hline
    ${\cal IP}_{8}$ & $4$ & $8$-ary Q-PAM & $N = 2(2^r - 1)$ & $(N + 2)/4$ & 3 &
    $\sqrt{N}$ \\
    \hline \hline
\end{tabular}
\end{center}
\end{table*}

\subsection{Prior Constructions in the Literature}

In this paper, keeping in mind the widespread usage of binary
digits to represent data, we restrict our attention to
low-correlation sequence families whose symbol-alphabet is a
subset of the complex numbers having size that is a power of $2$.

We do not consider sequences over real-valued alphabet such as the
BPSK $\{\pm 1\}$ alphabet or the PAM alphabet since apart from
their inherent ability to provide increased spectral efficiency,
the corresponding complex counterparts of these alphabets, namely
QPSK and QAM, offer better correlation performance in general. For
instance, for family sizes that are approximately equal to the
sequence period $N$, $\theta_{\max}$ for the best known BPSK and
QPSK sequence families is approximately given by $\sqrt{2N}$ and
$\sqrt{N}$ respectively \cite{Gold,BozHamKum}.

Table~\ref{table:others} provides a quick overview of some relevant
prior constructions:

\begin{itemize}

\item The quaternary sequence
family, Family ${\cal A}$ \cite{Sol,BozThesis,BozHamKum}, has the
same size as the family of Gold sequences \cite{Gold}, but smaller
value of $\theta_{\max}$.

\item Quaternary families $\{ {\cal S}(p) \}$, $p \geq 1$,
\cite{KumHelCalHam} are larger families with correspondingly larger
values of $\theta_{\max}$ and a member of these families, namely
Family ${\cal S}(2)$, appears in the W-CDMA standard \cite{3gpp} as
the short scrambling code.

\item In \cite{KumHelCal}, a Galois-ring analogue of the
Weil-Carlitz-Uchiyama (WCU) bound on exponential sums over finite
fields is derived and a general technique for constructing
low-correlation $2^m$-PSK sequences is presented that is based on
this bound.  In the table, the label WCU is used to refer to
sequence families constructed using this technique.

\item A $16$-QAM CDMA family ${\cal Q}_B$ is constructed in
\cite{Boz} by Bozta\c{s}.  We became aware of this construction
only much after the initial writing of this paper, see
\cite{AnaKum}.   As is the case with the sequence families
constructed here, Family ${\cal Q}_B$ is built up of quaternary
sequences drawn from QPSK Family ${\cal A}$ and is described in
greater detail in Section~\ref{sec:Boztas}.
\end{itemize}

\subsection{Notation and Nomenclature} \label{sec:notation_term}

Unless otherwise specified, the word sequence appearing in this
paper, will be a reference to a spreading sequence.

We interchangeably use the terms $\mathbb{Z}_4$ sequences (i.e.,
sequences over the integers $\pmod{4}$), $4$-phase sequences or
$4$-QAM sequences (sequences over $\{\pm 1 \pm \imath\}$) to refer
to quaternary sequences in this paper.

The constellation-size parameter $M$ will always be a power of
$2$, and more specifically be given by $M=2^m$.  With the
exception of the interleaved sequence families which have double
the period, the period of every sequence family described here is
of the form $N=2^r-1$.  We set $q=2^r$ keeping in mind that the
finite field of size $2^r$ plays a major part in the construction
of sequences having this period.

In many of the constructions presented here, each user is assigned a
subset of spreading sequences to choose from. Thus such a sequence
Family ${\cal X}$ is more accurately described as a collection of
subsets of sequences.  Despite this, to simplify presentation, we
will often refer to a selected sequence from one of the subsets
assigned to a user as either that user's spreading sequence or else
as a sequence belonging to Family ${\cal X}$.

\vspace*{0.1in}

\begin{defn} We shall say that a sequence over a
symbol alphabet ${\cal K}$ of size $K$ and of period $N$ is {\em
approximately balanced} if the number $\mu_j$ of times each symbol
$\lambda_j \in {\cal K}$ appears in one period of the sequence,
satisfies a bound of the type \bean
    \left| \mu_j -\frac{N}{K} \right| & \leq & O(\sqrt{N}) .
\eean
\end{defn}

\vspace*{0.1in}

\begin{defn} By the {\em data rate} of a sequence family, we will mean
the maximum number $\nu$ of bits that can be modulated onto each
spreading sequence within the family, while leaving the maximum
correlation parameter $\theta_{\max}$ undisturbed.  Thus the
sequence family would be capable of transferring $\nu$ bits per
period of the spreading sequence.
\end{defn}

\subsection{Principal Results} \label{sec:results}


The $M^2$-QAM constellation is the set \beq
    \{ a + i \, b \ \mid -M + 1 \leq a, b \leq M - 1 \ , \ \ a, b \ \mbox{odd}
    \}.\label{eq:QAM_constellation_natural}
\eeq When $M=2^m$, this constellation can alternately be described
as \cite{LuKum,TarSad} \beq
    \left\{ \sqrt{2\imath} \left( \left. \sum_{k = 0}^{m - 1} \, 2^k \, \imath^{a_k}
    \right)
    \ \right| \ a_k \in \mathbb{Z}_4 \right\} ,
    \label{eq:equivalent_QAM_expression}
\eeq where by $\sqrt{2 \imath}$ we mean the element $(1 +
\imath)=\sqrt{2} \exp(\frac{\imath 2 \pi}{8})$.

The class of Q-PAM constellations considered in this paper is the
subset of the $M^2$-QAM constellation of size $2M=2^{m+1}$ having
representation \beq
    \left\{ \sqrt{2 \imath} \left( \left. \imath^{a_0}\ + \ \sum_{k = 1}^{m - 1} \, 2^k \, (\imath)^{a_0+2a_k}
    \right)
    \ \right| \ \begin{array}{c} a_0 \in \mathbb{Z}_4, \\ a_k \in \mathbb{Z}_2, k \geq 1 \end{array} \right\} .
    \label{eq:AM_PSK_constellation}
\eeq

These representations suggest that quaternary sequences be used in
the construction of low correlation sequences over these
constellations.

We present five sequence families which adopt this approach, three
over the QAM constellation and two over a quadrature-PAM (Q-PAM)
subset of the QAM constellation.  All sequence families permit
data modulation at a rate higher than the $2$ bits per sequence
period permitted by the use of QPSK spreading sequences. Our
initial efforts were directed only at the QAM constellation until
we inadvertently discovered that correlation properties could be
improved by restricting the alphabet to the Q-PAM constellation,
while still retaining the higher data rate property.

All the sequence families constructed here also permit variable-rate
signalling. By this we mean that users can adjust their data rate by
switching to a spreading sequence over a constellation of the same
type, but of smaller or larger size. Interestingly, as we show, even
in the presence of variable-rate signalling, the amount of
interference experienced by a user remains essentially unchanged.

In all the constructions, the size of the sequence family is of the
order of $\frac{N}{\log_2(M)}$ where $M$ is the square root of the
size of the QAM constellation in the case of a QAM family, and
one-half the size of the Q-PAM constellation in the case of a family
over the Q-PAM signalling alphabet.

\subsubsection{Family ${\cal CQ}_{M^2}$}

This may be regarded as the canonical QAM sequence family
construction. In this family, each sequence $\{s(t)\}$ is of the
form \bea s(t) & = & \sqrt{2\imath}\sum_{k=0}^{m-1}2^k
\imath^{u_k(t+\tau_k)} \label{eq:expn_terms_Fam_A} \eea where the
sequences $\{u_k(t)\}$ are drawn from Family ${\cal A}$. The phases
$\tau_k$ are used to ensure that the sequence is approximately
balanced over its symbol alphabet.

Family ${\cal CQ}_{M^2}$ has period $N=2^r-1$ and normalized
maximum-correlation parameter $\overline{\theta}_{\max}$ bounded
above by $ \lesssim a \, \sqrt{N}$, where $a$ ranges from $1.8$ in
the $16$-QAM case to $3.0$ for large $M$.  The data rate in a CDMA
setting is $2m$.  This number can however, be increased to $3m$ bits
of data by halving the size of the sequence family and assigning
double the number of quaternary sequences to each user. Note that in
comparison, if one were to attempt to increase data rate with a QPSK
sequence family by assigning multiple sequences to each user, then
to increase the data rate by $m$, one would have to assign $2^m$
sequences to each user, thereby reducing the size of the family by a
factor of $2^m$.

Parameters of Family ${\cal CQ}_{M^2}$ are presented in
Table~\ref{table:seq_qam_nom}.   The corresponding parameters for
the case when the data rate is increased to $3m$ are presented in
Table~\ref{table:seq_qam_nom_data}.

The construction used to construct ${\cal CQ}_{M^2}$ is easily
extended to produce larger sequence families and an example is
provided in Section~\ref{sec:larger_canonical_fam}.

\subsubsection{Family ${\cal SQ}_{M^2}$} This family which we call, the ``selected''
family, has a lower value of $\theta_{\max}$ but permits only
$(m+1)$-bit data modulation. The sequences in this family also can
be described by equation \eqref{eq:expn_terms_Fam_A}.  The lower
value of $\theta_{\max}$ is made possible here by a judicious
selection of the component quaternary sequences
$\{u_k(t)\}_{k=0}^{m-1}$.  Parameters of this construction are
presented in Table~\ref{table:seq_qam}.

\subsubsection{Family ${\cal IQ}_{16}$} The $16$-QAM sequence
Family ${\cal IQ}_{16}$ has the best correlation properties, having
the lowest bound $\theta_{\max} \lesssim \sqrt{2} \, \sqrt{N}$
amongst all the $16$-QAM sequence families constructed in this
paper.  The family has data rate $3$ and is constructed using
sequence interleaving. Relevant parameters of the family are listed
in Table \ref{tab:qam_16}.

The remaining two families are over the Q-PAM constellation of
size $2M$.

\subsubsection{Family ${\cal P}_{2M}$ } This family has a lower
value of $\theta_{\max}$ in comparison with selected QAM Family
${\cal SQ}_{M^2}$, while still maintaining a data rate of $(m+1)$.
 Lower correlation values are obtained by setting all $\tau_k=0$ in
\eqref{eq:expn_terms_Fam_A} followed by adopting the sequence
selection used to construct Family ${\cal SQ}_{M^2}$. Setting
$\tau_k=0$ results in a sequence over a Q-PAM constellation (see
Table~\ref{table:seq_am_psk} for parameters of this sequence
family).

\subsubsection{Family ${\cal IP}_8$} This construction combines features
of constructions described above, namely, all $\tau_k=0$, judicious
sequence selection and sequence interleaving.  It achieves a data
rate of $3$ and quite remarkably, achieves the Welch lower bound on
$\overline{\theta}_{\max}$ (see Table~\ref{table:opt} for
parameters).

\subsection{Outline of the Paper}

Section \ref{sec:back} provides background material relating to the
QAM constellation, to Galois rings, to quaternary Family ${\cal A}$
and $16$-QAM Family ${\cal Q}_B$. Sections~\ref{sec:nom_fam},\
\ref{sec:Fam_A_M^2} discuss the canonical and selected QAM sequence
families ${\cal CQ}_{M^2}$ and ${\cal SQ}_{M^2}$ respectively.

In Section~\ref{sec:qam_16}, constructions for $16$-QAM sequences
are discussed.  Family ${\cal SQ}_{16}$, is shown to have
correlation properties that improve upon those of Family ${\cal
Q}_B$.  A second sequence family, Family ${\cal IQ}_{16}$,
introduced in this section and obtained using sequence
interleaving, is shown to do even better.

Section~\ref{sec:am_psk} deals with Q-PAM families, the general
construction of Family ${\cal P}_{2M}$ as well as the specific
$8$-ary Q-PAM construction ${\cal IP}_8$ that achieves the Welch
bound with equality.  Most proofs have been moved to the Appendix
for the sake of clarity.

\section{Background} \label{sec:back}

\subsection{The $M^2$-QAM and $2M$-ary Q-PAM Constellations}

The equivalence between the two representations of the $M^2$-QAM
constellations contained in \eqref{eq:QAM_constellation_natural} and
\eqref{eq:equivalent_QAM_expression} follows from noting that an odd
number, $x$, in the range $[-M + 1, M - 1]$ can be uniquely
expressed as \beqn
    x = \sum_{k = 0}^{m - 1} 2^k (-1)^{x_k} \ , \ \ x_k \in \mathbb{F}_2
\eeqn and the relation \beqn
    (-1)^{x_i + x_j} + \imath (-1)^{x_j} = \sqrt{2 \imath} \, \imath^{x_i + 2 x_j}
    \ , \ \ x_i, x_j \in \mathbb{F}_2.
\eeqn   As noted in Section~\ref{sec:results}, the representation
in \eqref{eq:equivalent_QAM_expression} suggests that a sequence
over $M^2$-QAM can be constructed using a collection of sequences
over $\mathbb{Z}_4$, of size $m$, and we adopt this approach in
the paper.   We shall also construct sequences over the Q-PAM
constellation described by \eqref{eq:AM_PSK_constellation}.

\begin{figure}[!ht]
    \begin{center}
        \centerline{\epsfig{figure=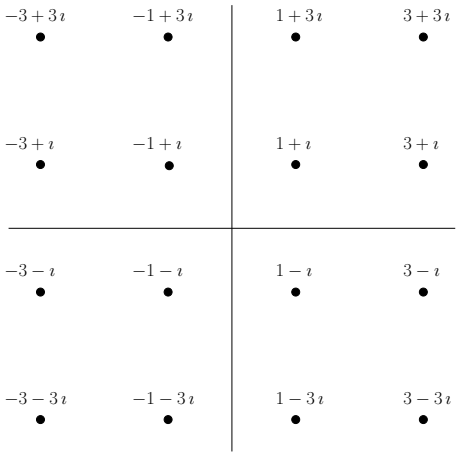,width=60mm}}
        \caption{{$16$-QAM Constellation \label{fig:16qam}}}
    \end{center}
\end{figure}

The $16$-QAM constellation~\cite{Boz,LuKum,RobTar} \beqn
    \left\{ \sqrt{2 \imath} \left( \imath^{a_0} + 2 \, \imath^{a_1} \right)
    \ \mid \ a_0, a_1 \in
    \mathbb{Z}_4 \right\}
\eeqn is shown in Fig. \ref{fig:16qam}.  It is easy to check that
the average energy of the constellation is $10$.

\begin{figure}[!ht]
    \begin{center}
        \centerline{\epsfig{figure=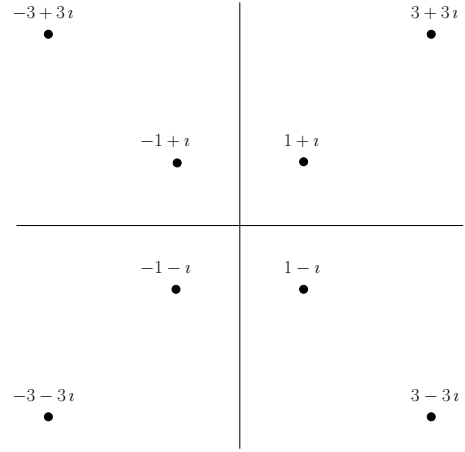,width=60mm}}
        \caption{{$8$-ary Q-PAM Constellation \label{fig:am_psk}}}
    \end{center}
\end{figure}

The $8$-ary Q-PAM constellation is a subset of the $16$-QAM
constellation given by \beqn
    \left\{ \sqrt{2 \imath} \left( \imath^{a_0} + 2 \, \imath^{a_0 + 2 a_1} \right)
    \ \mid \ a_0 \in \mathbb{Z}_4 \ , \ \ a_1 \in
    \mathbb{Z}_2 \right\},
\eeqn  (see Fig.~\ref{fig:am_psk}) and has the same average
energy.

\subsection{Galois Rings} \label{sec:math_prelim}

Let $\mathbb{Z}_n$ denote the ring of integers modulo $n.$  In this
paper our primary interest is in the ring $\mathbb{Z}_4 = \{ 0, 1,
2, 3\}.$

Galois rings \cite{Mac} are Galois extensions of the prime ring
$\mathbb{Z}_{p^n}$. $R \triangleq GR(4, r)$ will denote a Galois
extension of $\mathbb{Z}_4$ of degree $r$. $R$ is a commutative ring
with identity and contains a unique maximal ideal $M=2R$ generated
by the element $2$.  Such rings are called local rings.   The
quotient $R / M$ is isomorphic to $\mathbb{F}_q$, the finite field
with $q = 2^r$ elements.

As a multiplicative group, the set $R^{*}$ of units of $R$ has the
following structure: \beqn
    R^{*} \cong \mathbb{Z}_{2^{r}-1} \times \underbrace{\mathbb{F}_2 \times
    \mathbb{F}_2 \ldots \times \mathbb{F}_2}_{r \mbox{ times}}.
\eeqn Let $\xi$ be a generator for the multiplicative cyclic
subgroup isomorphic to $\mathbb{Z}_{2^{r} - 1}$ contained within
$R^{*}$.  Let $\mathcal{T}$ denote the set $\mathcal{T} = \{ 0, 1,
\xi, \ldots, \xi^{2^r - 2}\}$.  ${\cal T}$ is called the set of
Teichmueller representatives (of $\mathbb{F}_q$ in $R$).  It can be
shown that every element $z \in R$ can uniquely be expressed as
\beqn
    z = a + 2 b, \ \ a, b \in \mathcal{T}.
\eeqn This is often referred to as the ``$2$-adic expansion'' of
$z$.  Modulo-$2$ reduction of $z$ is denoted by $\overline{z}$.   It
can be shown that $\alpha = \overline{\xi}$ is a primitive element
in $\mathbb{F}_q$.

To every element $a \in \mathbb{F}_q$ there exists a unique element
$\hat{a}$ in $\mathcal{T}$ such that $\overline{\hat{a}}=a$.  The
element $\hat{a}$ is called the ``lift'' of $a$ in $R$.

\begin{note}
To simplify notation, in the sequel, we will often use the same
notation to refer to both the finite field element as well as its
lift belonging to the associated Teichmuler set ${\cal T}$.
\end{note}

The Frobenius automorphism $\sigma:R \to R$ is given by \beqn
    \sigma(z) = a^2 + 2 b^2
\eeqn and the trace map from $R$ to $\mathbb{Z}_4$ is defined as
\beqn
    T(z) = \sum_{k = 0}^{r - 1} \, \sigma^k(z) \ = \
    \sum_{k=0}^{r-1} (a^{2^k} + 2 b^{2^k}).
\eeqn Let $tr: \mathbb{F}_q \to \mathbb{F}_2$ denote the binary
trace function.  More details of Galois rings can be found in
\cite{Mac,HamKumCalSloSol,Sha,KumHelCal}.

\subsection{Modifications to the Maximum Correlation Parameter} \label{sec:seq_corr}

We make two changes to the maximum correlation parameter.  The
first change recognizes that when a user is assigned multiple
spreading sequences, a bank of correlators is used at the receiver
end and the autocorrelation between two such sequences at zero
shift does not interfere with the self-synchronization capability
of the family.   Accordingly, the maximum non-trivial correlation
magnitude of a sequence family is given the modified definition:
\beq
    \theta_{\max} := \max \left\{ \left| \theta_{s(j), \ s(k)}(\tau) \right|
    \left| \begin{array}{l}
    \mbox{either } s(j, t) , \, s(k, t) \\
    \mbox{have been assigned}\\
    \mbox{to distinct users or} \\
    \tau \neq 0
    \end{array} \right.\right\}.
    \label{eqref:theta_max}
\eeq

The second change arises from energy considerations.  To make a fair
comparison between QAM and PSK families, it is required that the
correlation magnitude be normalized to take into account the larger
energy of the QAM and Q-PAM sequence families.  We will use
$\overline{\theta}_{\max}$ to denote the maximum correlation
magnitude if the sequences have been normalized to have energy
$N$, and this will be used as the basis for comparison across signal
constellations.

\subsection{Family ${\cal A}$} \label{sec:Fam_A}

Family ${\cal A}$ is an asymptotically optimal family of quaternary
sequences (i.e., over $\mathbb{Z}_4$) discovered independently by
Sol\'e\cite{Sol} and Bozta\c{s}, Hammons and Kumar
\cite{BozThesis,BozHamKum}.  A detailed description of their
correlation properties appears in \cite{BozHamKum}.

Let $\{\gamma_j\}_{j=1}^{2^r}$ denote $2^r$ distinct elements in
$\cal T$, i.e., we have the alternate expression $\mathcal{T} = \{
\gamma_1, \gamma_2, \ldots, \gamma_{2^r} \}$.  There are $2^r + 1$
cyclically distinct sequences in Family $\cal A$, each of period
$N=2^r - 1$.  The following representation for sequences in Family
${\cal A}$ is used in this paper: \beqa
    s_j(t) & = & T([1 + 2 \, \gamma_j] \xi^t), \ \ \ 1 \leq j \leq 2^r,
    \nonumber \\
    s_{2^r + 1}(t) & = & 2 \, T(\xi^t). \label{eqref:def_FamA}
\eeqa

The maximum non-trivial correlation magnitude for Family ${\cal A}$
has the upper bound \beq
    \theta_{\max} \leq 1 + \sqrt{N + 1}. \label{eqref:cc_FamA}
\eeq  More details of the correlation properties of Family $\cal
A$ can be found in Appendix~\ref{app:closed_form_Fam_A}. Various
desirable properties such as near optimality with respect to
correlation, mathematical tractability and ease of generation,
make Family ${\cal A}$ a prime candidate for use as a building
block in constructing sequences over the $M^2$-QAM constellation.

\begin{note} In the present paper, we do not make
use of the presence of the ``binary'' sequence $\{s_{2^r+1}(t)\}$ as
our sequence constructions require each quaternary sequence employed
to take on all possible values over $\mathbb{Z}_4$.  For this
reason, we will treat Family ${\cal A}$ as if it were a family
composed of $q=2^r$ cyclically-distinct sequences.   A similar
comment applies in Section~\ref{sec:larger_canonical_fam} where we
make use of quaternary sequence Family ${\cal S}(1)$.
\end{note}

\subsection{The $16$-QAM Sequence Family Constructed by Bozta\c{s}} \label{sec:Boztas}

In \cite{Boz}, Bozta\c{s} considers the family of sequences \bean
    {\cal Q}_{B} & = & \left\{ \alpha u_i(t) \ + \ \beta v_i(t) \mid 1
    \leq i \leq 2^r - 2 \ \right\}
\eean where $\alpha, \beta $ are positive real numbers.  The
sequences $\{u_i(t)\}$ are defined as follows.   We adopt the
notation introduced in Section~\ref{sec:math_prelim} relating to a
Galois ring $R = GR(4,r)$ of size $4^r$.  Let the elements
$\delta_i, \gamma_i$ be any selection satisfying \bean
    \left\{ \delta_1, \delta_2, \cdots, \delta_{2^{r-1}-1} \right\}
    \bigcup \left\{ \gamma_1, \gamma_2, \cdots, \gamma_{2^{r-1}-1} \right\}
    \\ \ = \  \left\{(1-\xi), (1-\xi^2), \cdots, (1-\xi^{2^r-2})
    \right\}
\eean where $\xi$ is, as in Section~\ref{sec:math_prelim}, a
generator for the multiplicative cyclic subgroup isomorphic to
$\mathbb{Z}_{2^{r} - 1}$ contained within $R^{*}$.  Then the
sequences $\{u_i(t)\}, \{v_i(t)\}$ are defined by \bean
    u_i(t) & = & \imath^{T(\delta_i \xi^t)} \\
    v_i(t) & = & \imath^{T(\gamma_i \xi^t)} .
\eean The resulting sequence family ${\cal Q}_{B}$ has in general, a
constellation of size $16$.  Although not explicitly pointed out in
\cite{Boz}, by setting $\alpha=1$ and $\beta=2$, one recovers a
rotated version of the 16-QAM constellation: \beqn
    \{ a + i \, b \ \mid -3  \leq a, b \leq 3 \ , \ \ a, b \ \mbox{odd}  \}.
\eeqn In Table~\ref{table:others}, we have listed parameters of the
family ${\cal Q}_B$ obtained by selecting $\alpha=\sqrt{2 \imath}$,
$\beta=2 \alpha$.   While some discussion of the correlation
properties of this sequence family is presented in \cite{Boz}, the
value of $\theta_{\max}$ for Family ${\cal Q}_B$ listed in
Table~\ref{table:others} is derived from the results of the present
paper.

\section{Canonical families of sequences over $M^2$-QAM constellation} \label{sec:nom_fam}

The equivalent expression for the QAM constellation given in
\eqref{eq:equivalent_QAM_expression} suggests that a family of low
correlation sequences can be constructed using, as building
blocks, elements of Family A as follows: \bean
    s_p(t) & = & \sum_{k=0}^{m-1} 2^k \imath^{T(a_{p, k} \xi^t)} \ ,
\eean where the coefficients $\{ a_{p, k} \}$ are drawn from
$GR(4,r)$. When one considers the crosscorrelation between two
sequences $\{ s_{p}(t) \}$ and $\{ s_{q}(t) \}$ having the above
form, one quickly realizes that in order to keep correlation values
small, no two sequences $\{\imath^{T(a_{p, k} \xi^t)}\}$,
$\{\imath^{T(a_{q, l} \xi^t)}\}$ should be cyclic shifts of one
another.  This requirement can equivalently be expressed in the form
\bean
    a_{p, k} \xi^{\tau} & \neq & a_{q, l}
\eean for any value of the cyclic shift parameter $\tau$, whenever
either $p \neq q$ (signifying different users) or whenever $k \neq
l$, $0 \leq k,l \leq m-1$.

Let us impose the second requirement on the QAM sequence family
that every sequence in the family should be approximately balanced
as defined in \ref{sec:notation_term}.  This requires that the
number of solutions to the simultaneous equations: \bean
    T(a_{p, k} x) & = & \nu_k \ , \ \ k = 0,1, \ldots, (m-1)
\eean be approximately equal for all $m$-tuples $\nu \ = \ (\nu_0,
\nu_1, \ldots, \nu_{m-1})$ in $\mathbb{Z}_4^m$ as $x$ varies over
all of $\cal T$.

\vspace*{0.1in}

\begin{lem} \label{lem:bal_gen}
The sequence \bean
    s_p(t) & = & \sum_{k = 0}^{m - 1} 2^k \imath^{T(a_{p,k} \xi^t)}
\eean is approximately balanced over the $4^m$-QAM alphabet if the
coefficients $a_{p, k}$ are linearly independent over
$\mathbb{Z}_4$, i.e., \bean
    \sum_{k = 0}^{m - 1} \omega_k a_{p, k} & \neq & 0
\eean for any choice $\{\omega_k \in \mathbb{Z}_4 \}_{k=0}^{m-1}$
of coefficients, where at least one of the $\omega_k$'s is
non-zero.
\end{lem}

\vspace*{0.1in}

\begin{proof} Please see Appendix~\ref{app:bal_gen_pf}. \end{proof}

\vspace*{0.1in}

From the above discussion we arrive at the twin conditions \beq
    a_{p, k} \xi^{\tau} \ \neq \ a_{q, l}
     \label{eq:condition_1}
\eeq whenever either $p \neq q$ or $k \neq l$ and \beq
    \sum_{k = 0}^{m - 1} \omega_k a_{p,k} \ \neq \ 0  \label{eq:condition_2}
\eeq for any non-zero coefficient set $\{\omega_k\}_{k = 0}^{m-1}$;
which we will respectively term as the {\em cyclic distinctness} and
{\em linear independence} conditions to be satisfied by the
coefficients $\{a_{p,k}\}$.

One means of constructing coefficient sets $\{ a_{p,k} \}$
satisfying the twin conditions in \eqref{eq:condition_1} and
\eqref{eq:condition_2} is described below.

\vspace*{0.2in}

Let $\{ \tau_0 = 0, \tau_1, \tau_2, \ldots, \tau_{m-1} \}$ be
integers $0 \leq \tau_i \leq 2^r-2$, such that $\{ \alpha^{\tau_0} =
1, \alpha^{\tau_1}, \alpha^{\tau_2}, \ldots, \alpha^{\tau_{m-1}} \}$
form a linearly independent set.   Let the elements of the
Teichmuller ${\cal T}$ set be divided into disjoint (ordered)
subsets, each of size $m$, of the form $g = (g_0, g_1, \ldots, g_{m
- 1})$.   Let $G$ refer to the collection of all such $g$'s. Note
that \bean \mid G \mid  & = & \left\lfloor \frac{q}{m} \right\rfloor
\ = \ \left\lfloor \frac{N+1}{m} \right\rfloor. \eean Set \bean
    a_{g,k} & = & (1 + 2 g_k)\xi^{\tau_k} .
\eean It is straightforward to verify that the coefficients
$a_{g,k}$ satisfy the cyclic distinctness requirement.  To see that
the linear independence requirement is also met, note that \bean
    \sum_k \omega_k a_{g,k} & = & 0 \label{eq:lin_ind}
\eean implies \bean
    \sum_k \omega_k \xi^{\tau_k} & = & 0 \pmod{2}
\eean which is not possible by the choice of $\{\tau_k\}$ unless
all the $\omega_k \in \{0,2\}$.   But this possibility can also be
dismissed using a similar argument.

This leads to the construction of a family of $M^2$-QAM sequences
which we shall term the {\em canonical} construction and denote by
Family ${\cal CQ}_{M^2}$ .

Let $\kappa = (\kappa_0, \kappa_1, \ldots, \kappa_{m - 1}) \in
\mathbb{Z}_4 \times \mathbb{F}_2^{m - 1}$.   A mathematical
expression for Family ${\cal CQ}_{M^2}$ is provided below. \beqa
    {\cal CQ}_{M^2} & = & \left\{
    \left. \left\{ s(g, \kappa, t) \left| \begin{array}{l}
    \kappa \in \mathbb{Z}_4^m
    \end{array} \right\} \right. \right|
    g \in G \right\}. \nonumber\\ \label{eqref:QN_M^2_defn}
\eeqa Each user is thus assigned the set \beqn
    \left\{ s(g, \kappa, t) \ | \
    \kappa \in \mathbb{Z}_4^m \right\}
\eeqn of sequences with the $\kappa$-th sequence given by \beqan
    s(g, \kappa, t) & = & \sqrt{2 \imath} \left( \sum_{k = 0}^{m - 1}
    2^k \, \imath^{u_k(t)} \imath^{\kappa_k} \right)
\eeqan where \bea
    u_k(t) & = & T([1 + 2 g_k]\xi^{t + \tau_k}), \nonumber\\
    & & \hspace{1in} k = 0, 2, \ldots, m - 1.
    \label{eq:family_A_CQ_component}
\eea

The main properties of Family ${\cal CQ}_{M^2}$ are summarized in
the following theorem:

\vspace{0.1in}

\bthm \label{thm:main_cq} Let $m \geq 2$ be a positive integer and
let ${\cal CQ}_{M^2}$ be the family of sequences over the
$M^2$-QAM constellation defined in (\ref{eqref:QN_M^2_defn}).
Then,

\ben

\item \label{thm:main_cq_prd} All sequences in ${\cal CQ}_{M^2}$
have period $N = 2^r - 1$.

\item \label{thm:main_cq_energy} For large values of $N$, the
energy of the sequences in the family is given by \beqn
    {\cal E} \approx \frac{2}{3} \, (M^2 - 1) \, N, \eeqn
(which is what one would expect if the average energy of a symbol
across the constellation were equal to the average symbol energy
across one period of the sequence).

\item \label{thm:main_cq_corr}  The maximum correlation parameter
of the family can be bounded  as \beqan
    \theta_{max}  \ \lesssim \
    2 \left( 2^m - 1 \right)^2 \sqrt{N + 1}.
\eeqan

For large values of $M$ and $N$, the normalized maximum
correlation parameter of the family can be bounded as \beqn
    \overline{\theta}_{\max} \lesssim 3 \, \sqrt{N}.
\eeqn

\item \label{thm:main_cq_sz_mod} Family ${\cal CQ}_{M^2}$ can
support $\lfloor(N + 1)/m \rfloor$ distinct users.

\item \label{thm:main_cq_data} Each user can transmit $2 m$ bits of information per
sequence period.

\item \label{thm:main_cq_ed} The normalized minimum squared
Euclidean distance between all sequences assigned to a user is
given by \beqn
    \overline{d}_{\min}^{\, 2} \approx \frac{6}{M^2 - 1} \, N.
\eeqn

\item \label{thm:main_cq_bal} The sequences in Family ${\cal CQ}_{M^2}$ are
approximately balanced.

\een \ethm

\vspace{0.1in}

\bpf Property (\ref{thm:main_cq_prd}) follows from the periodicity
properties of Family ${\cal A}$ sequences. Properties
(\ref{thm:main_cq_energy}) and (\ref{thm:main_cq_corr}) follow
from the correlation properties of Family ${\cal A}$ sequences and
the derivation may be found in
Appendices~\ref{app:main_cq_energy_pf} and
\ref{app:main_cq_corr_pf} respectively.

In order to prove property (\ref{thm:main_cq_sz_mod}), we note that
each user is assigned $m$ cyclically distinct sequences from Family
${\cal A}$, namely the sequence set \bean
    \{T([1 + 2g_{k}]\xi^{t+\tau_k}\}_{k=0}^{m-1}.
\eean  Since there are $q$ possible choices for $g_{k}$ in ${\cal
T}$, it follows that the maximum number of users that can be
supported is given by $\lfloor 2^r/m \rfloor = \lfloor (N+1)/m
\rfloor $.  Property (\ref{thm:main_cq_data}) follows from the
definition of the sequences.  The symbols $\{ \kappa_k \}_{k = 0}^{m
- 1}$ are the $m$ information-bearing symbols.  Property
(\ref{thm:main_cq_ed}) is concerned with the Euclidean-distance of
the sequences assigned to a user and is proved in
Appendix~\ref{app:main_cq_ed_pf}. Property (\ref{thm:main_cq_bal})
follows from our earlier arguments. \epf

\vspace*{0.1in}

\begin{note} \label{not:energy}
An examination of the proof of Property~(\ref{thm:main_cq_energy})
will reveal that the result is valid for any $M^2$-QAM, ($M^2=4^m$),
sequence $\{s(t)\}$, given by an expression of the form: \bean
    s(t) & = & \sqrt{2\imath} \sum_{k=0}^{m-1} 2^k \imath^{u_k(t)}
\eean where the component quaternary sequences $\{u_k(t)\}$, are
distinct elements of Family ${\cal A}$.
\end{note}

\subsection{Variable-Rate Signalling}

The asynchronous nature of the reverse link (mobile to base station)
in a CDMA system makes it difficult to accommodate users having
differing data-rate requirements i.e., users who wish to communicate
a different number of bits of data per sequence period. It
precludes, for example, the use of
orthogonal-variable-spreading-factor (OVSF) channelization (Walsh)
codes that are part of the WCDMA standard.

One of the advantages of the structure of the sequences in Family
${\cal CQ}_{M^2}$ (and others presented here) is that it is possible
to place an upper bound on the crosscorrelation of sequences over
QAM constellations of different size, thereby enabling variable-rate
signalling on the reverse link.

Enabling variable-rate signalling in the case of Family ${\cal
CQ}_{M^2}$ is fairly straightforward as we shall see.  One first
partitions the entire finite field into subsets, and the subsets
will typically be of different sizes. The elements in each subset
are then ordered in some arbitrary fashion, and if $\{ g_k
\}_{k=0}^{m-1} \subseteq \mathbb{F}_q$ is the ordered subset, then
this subset is associated with the $4^m$-QAM sequence \bean
    \sqrt{2\imath} \sum_{k=0}^{m-1} 2^k \imath^{T([1+2
    g_k]\xi^{t+\tau_k})}.
\eean

Thus every partition of the elements of the Teichmuller set
corresponds to an assignment of variable rates to the users, with
the number of users equal to the number of subsets in the
partition.   It follows that we can support $n_i, i=1,2,\cdots,p$
users with constellations of size $4^{m_i}$ iff \bean \sum_{k=1}^p
n_i m_i & \leq & \mid {\cal T } \mid \ = \ q.  \eean

If two users have been assigned tuples enabling them to transmit
sequences from Families ${\cal CQ}_{M_1^2}$ and ${\cal
CQ}_{M_2^2}$, with $M_1>M_2$, then the user assigned sequences
from  Family ${\cal CQ}_{M_2^2}$ will experience a marginally
increased amount of interference from the user assigned sequences
from Family ${\cal CQ}_{M_1^2}$.  The reverse is true in the case
of the interference experienced by the user having the larger
constellation.  This is based on the bounds on normalized
crosscorrelation~\footnote{The reader can readily verify that
normalized crosscorrelation is the right measure to employ here.}
derived in Appendix~\ref{app:var_pf}. The marginal change is by a
factor of \beqn
    \sqrt{\frac{(M_1-1)(M_2+1)}{(M_1+1)(M_2-1)}} .
\eeqn It is this essentially-unchanged level of interference that
enables variable-rate signalling.

\subsection{Euclidean Distance Comparison with $M^2$-PSK Constellation}
\label{sec:comp}

Each sequence belonging to Family ${\cal CQ}_{M^2}$ can be modulated
by $\log{M^2}=2m$ data bits.  An alternative means of transporting
$2m$ data bits per period of spreading sequence, is to use a QPSK
code sequence family and then use $M^2$-ary phase data modulation
which corresponds to multiplication of the code sequence by a
complex symbol drawn from the set  \beqn
    \left\{ \exp\left(\imath \frac{2 \, \pi \, a}{M^2}\right) ,\ \ \
    a \in \mathbb{Z}_{M^2}\right\}.
\eeqn

We compare the two schemes in terms of the minimum Euclidean
distance between the same code sequence when modulated by two
different $2m$-tuples of data.  In the case of $M^2$-ary PSK
modulation, the minimum squared Euclidean distance between two
distinct modulations of a sequence over $M^2$-PSK can be shown to
be given by \bean 2 \, \left(1 - \cos \left( {\frac{2 \,
\pi}{M^2}} \right) \right) \, N, \eean where $N$ is the period of
the code sequence.  For large $M$ the right hand side can be
approximated by \bean
    2 \, \left(1 - \cos \left( {\frac{2 \, \pi}{M^2}} \right) \right) N
    & \approx &
    2\left(1 - \left(1-\frac{(\frac{2\pi}{M^2})^2}{2!}\right)\right))N \\
    & = & \frac{4 \pi^2}{M^4}N.
\eean

In comparison, the normalized minimum squared Euclidean distance
between two sequences assigned to a user in Family ${\cal
CQ}_{M^2}$ is given by \bean \frac{6}{(M^2 - 1)} N, \eean and it
is clear from this that Family ${\cal CQ}_{M^2}$ has significantly
larger separation between different data sets which makes for
increased reliability.

\subsection{Further Increasing the Data Rate}

In the present construction, Family ${\cal CQ}_{M^2}$ is a family of
$\lfloor \frac{N+1}{m} \rfloor$ sequences in which each user can
transmit $2m$ bits of data.  The sequence $\{s(g,\kappa,t)\}$ of
each user is built up of $m$ quaternary sequences $\{u_k(t) \}_{k =
0}^{m - 1}$ drawn from Family ${\cal A}$ and is of the form \bean
    s(g,\kappa,t) & = &
    \sum_{k=0}^{m-1} 2^k \imath^{u_k(t)+\kappa_k}
\eean where \beqn
    u_k(t) \ = \ T([1+2g_k]\xi^{t + \tau_k}) .
\eeqn

Suppose, we were to assign additional $m$ sequences
$\{v_k(t)\}_{k=0}^{m-1}$ from Family $\cal A$ to each user where
\bean
    v_k(t) & = & T([1 + 2h_k]\xi^{t + \tau_k}) .
\eean This would, on the one hand, reduce the family size by a
factor of $2$ to $\lfloor \frac{N+1}{2m} \rfloor$. On the other
hand, this would enable each user to transmit an additional $m$
bits of data per period of the code sequence.  The user could
simply select between the pair $\{u_k(t)\}$ and $\{v_k(t)\}$ for
the $k$-th component sequence.  There is no penalty to be paid in
terms of increased correlations since, as can easily be verified,
the maximum normalized correlation magnitude bound remains
unchanged. Decorrelation at the receiver end can be accomplished
with the aid of $2m$ decorrelators in place of the $m$ previously
needed.

Note that this feature is peculiar to the structure of the
signalling set used here.  If one were to attempt something
similar in conjunction with a QPSK sequence family, then in order
to send an additional $m$ data bits, one would have to assign each
user an additional $2^{m}-1$ code sequences and employ $2^m-1$
additional de-correlators at the receiver!

\subsection{Compatibility with Quaternary Sequence Families}

Being built up of quaternary sequences gives Family ${\cal
CQ}_{M^2}$ the added advantage of being compatible with QPSK
Families ${\cal S}(p)$ in the sense that the value of maximum
correlation magnitude is increased only slightly if one enlarges
Family ${\cal CQ}_{M^2}$ to include quaternary sequences drawn from
${\cal S}(p) \setminus {\cal A}$.  We omit the details.

\subsection{Larger Canonical Families over the QAM Alphabet}
\label{sec:larger_canonical_fam}

The canonical sequence family ${\cal CQ}_{M^2}$ described in
Theorem~\ref{thm:main_cq} was based on the use of Family ${\cal A}$
as the source for the component quaternary sequences $\{u_k(t)\}$
(see \eqref{eq:family_A_CQ_component}).  The construction extends
easily to the case when the component sequences are drawn from any
low-correlation quaternary family. In particular, one could
construct larger, low-correlation $M^2$-QAM families from the large
collection of low-correlation WCU quaternary sequence families (see
Table~\ref{table:others}).  We illustrate by considering the case
when Family ${\cal A}$ is replaced by quaternary sequence family
${\cal S}(1)$ and leave the details in the other cases to the
reader.   We will use the notation ${\cal CQ}_{M^2}({\cal S}(1))$ to
describe this sequence family.  Under this notation, ${\cal
CQ}_{M^2}$ is shorthand for Family ${\cal CQ}_{M^2}({\cal A})$.

Family ${\cal S}(1)$ contains $q^2$ cyclically distinct sequence
families.  Let $P=\lfloor \frac{q^2}{m}\rfloor $ and let a subset of
Family ${\cal S}(1)$ of size $Pm$ be selected.  Only this subset
will be used in the construction.   Then it can be shown that this
collection of $Pm$ cyclically-distinct sequences can be placed into
an array of size $P \times m$ in which the $(p,k)$-th element is of
the form \bean
    T([1+2g_{p,k}]\xi^{t+\tau_k} + 2h_{p,k}\xi^{3t}).
\eean   Let $\kappa = (\kappa_0, \kappa_1, \ldots, \kappa_{m - 1})
\in \mathbb{Z}_4^m$.   The signal of the $p$-th user, $1 \leq p \leq
P$, is then given by \bean
    s(p,\kappa,t) & = & \sqrt{2 \imath}
    \left( \sum_{k=0}^{m-1} 2^k \imath^{u_k(t)} \imath^{\kappa_k}
    \right)
\eean where \bean
    u_k(t) & = & T([1+2g_{p,k}]\xi^{t+\tau_k} +
    2h_{p,k}\xi^{3t}) .
\eean Then, Family ${\cal CQ}_{M^2}({\cal S}(1))$ is the collection
of sequences \bea
    {\cal CQ}_{M^2}({\cal S}(1)) 
    \ = \  \left\{ \{ s(p,\kappa,t) \mid \kappa \in \mathbb{Z}_4^m
    \} \mid 1 \leq p \leq P \right\}. \label{eq:CQ_S1_M^2_defn}
\eea

The main properties of Family ${\cal CQ}_{M^2}({\cal S}(1))$ are
summarized in the following theorem:

\vspace{0.1in}

\bthm \label{thm:main_cq_s} Let $m \geq 2$ be a positive integer and
let ${\cal CQ}_{M^2}({\cal S}(1))$ be the family of sequences over
the $M^2$-QAM constellation defined in \eqref{eq:CQ_S1_M^2_defn}.
Then,

\ben

\item \label{thm:main_cq_s_prd} All sequences in ${\cal
CQ}_{M^2}({\cal S}(1))$ have period $N = 2^r - 1$.

\item \label{thm:main_cq_s_energy} For large values of $N$, the
energy of the sequences in the family is given by \beqn
    {\cal E} \approx \frac{2}{3} \, (M^2 - 1) \, N. \eeqn

\item \label{thm:main_cq_s_corr}  The maximum correlation parameter
of the family can be bounded  as \beqan
    \theta_{max}  \ \lesssim \
    4 \left( 2^m - 1 \right)^2 \sqrt{N + 1}.
\eeqan

For large values of $M$ and $N$, the normalized maximum
correlation parameter of the family can be bounded as \beqn
    \overline{\theta}_{\max} \lesssim 6 \, \sqrt{N}.
\eeqn

\item \label{thm:main_cq_s_sz_mod} Family ${\cal CQ}_{M^2}({\cal
S}(1))$ can support $\lfloor(q^2)/m \rfloor$ distinct users.

\item \label{thm:main_cq_s_data} Each user can transmit $2 m$ bits
of information per sequence period.

\item \label{thm:main_cq_s_ed} The normalized minimum squared
Euclidean distance between all sequences assigned to a user is
given by \beqn
    \overline{d}_{\min}^{\, 2} \approx \frac{6}{M^2 - 1} \, N.
\eeqn

\een \ethm

\vspace*{0.1in}

\begin{proof}
The proof is along the same lines as used to prove the properties
of Family ${\cal CQ}_{M^2}$.  The principal difference is that
$\theta_{\max}({\cal S}(1)) \leq 2 \sqrt{N+1}$ in place of
$\theta_{\max}({\cal A}) \leq  \sqrt{N+1}$, see
\cite{KumHelCal},\cite{KumHelCalHam}.
\end{proof}


\section{A ``Selected'' Construction of Sequences Over $M^2$-QAM } \label{sec:Fam_A_M^2}

We now introduce a second family, Family ${\cal SQ}_{M^2}$, of
sequences over the $M^2$-QAM constellation having a lower value of
normalized correlation parameter $\overline{\theta}_{\max}$, and
twice the squared-Euclidean distance between different data
modulations of the same spreading sequence.  As against this, Family
${\cal SQ}_{M^2}$ permits users to transmit only $(m + 1)$ bits of
data per sequence period in place of the $2m$ bits allowed by Family
${\cal CQ}_{M^2}$.

Lower correlation values are achieved by judicious selection of the
component quaternary sequences constituting a QAM-sequence.

\subsection{Definition of Family ${\cal SQ}_{M^2}$} \label{sec:Fam_PQ_M^2_def}

Let $\{\delta_0=0,\delta_1, \delta_2, \ldots, \delta_{m - 1}\}$ be
elements from $\mathbb{F}_q$ such that $tr(\delta_k) = 1, \forall \
k \geq 1$.  Set \bean
    H & = & \{\delta_0, \delta_1 \cdots, \delta_{m-1} \}.
\eean Let $G =\{g_k\}$ be the largest subset of $\mathbb{F}_q$
having the property that \bea
    g_k + \delta_p & \neq & g_l + \delta_q \ , \ \
    g_k, g_l \in G \ , \ \ \delta_p, \delta_q \in H \ ,
    \label{eq:g_delta_sums_are_distinct}
\eea unless $g_k = g_l$ and $\delta_p = \delta_q$.  Then the
corresponding Gilbert-Varshamov and Hamming bounds on the size of
$G$ are given by \bea
    \frac{2^{r}}{1+\binom{m-1}{1}+\binom{m-1}{2}} \leq
    & \mid G \mid & \leq \frac{2^r}{{1+\binom{m-1}{1}}} \ .
    \label{eq:Hamming_GV_bounds}
\eea

\subsubsection{A Subspace-Based Construction for the $H$ and $G$}

Given constellation parameter $m$, let $2^l$ denote the smallest
power of $2$ greater than $(m-1)$, i.e., $l$ is defined by \bea
    2^{l-1} < (m - 1) \leq 2^l . \label{eq:m_1_and_powers_of_2}
\eea
For reasons that will shortly become clear, we will refer to the
integer $l$ as the {\em subspace-size exponent} (sse) associated
with the {\em constellation parameter} (c-p) $m$. Thus $l$ will
lie in the range $0 \leq l \leq (r-1)$.  Let $\mu$ denote the
function that, given c-p $m$ in the range $ 1\leq m \leq
2^{r-1}+1$, maps $m$ to the corresponding sse $l$ given above,
i.e., \bean
    \mu(m) & = & l.
\eean

Treating $\mathbb{F}_q$ as a vector space over $\mathbb{F}_2$ of
dimension $r$, let $W_{r-1}$ denote the subspace of $\mathbb{F}_q$
of dimension $(r-1)$ corresponding to the elements of trace $=0$.
Let $W_l$ denote a subspace of $W_{r-1}$ having dimension $l$. Let
$\zeta$ be an element in $\mathbb{F}_q$ having trace $1$ and let
$V_l$ denote the subspace
 \bean
    V_l & = & W_l \cup \{W_l+\zeta\}
\eean of size $2^{l+1}$.    Noting that every element in the coset
$W_l+\zeta$ of $W_l$ has trace $1$, we select as the elements
$\{\delta_k\}_{k=1}^{m-1}$ to be used in the construction of
Family ${\cal SQ}_{M^2}$, an arbitrary collection of $(m-1) \leq
2^l$ elements selected from the set $W_l+\zeta$.

Next, we partition $W_{r-1}$ into the
 $2^{r-l-1}$ cosets $W_l+g$ of $W_l$.  With
each coset, we associate a distinct user.  To this user, we assign
the coefficient set \bean
    \{g, g+\delta_1, g+\delta_2,\ldots, g+\delta_{l}\}.
\eean The coefficients $\{g+\delta_k\}_{k=1}^{m-1}$ belong to the
coset $W_l+(g+\zeta)$ of $W_l$.  Thus in general, each user is
assigned $m$ coefficients, with one coefficient $g$, belonging to
the coset $W_l+g$ of $W_l$ lying in $W_{r-1}$ and the remaining
drawn from the coset $W_l+g+\zeta$ of $W_l$.   Since $V_l=W_l \cup
(W_l+\zeta)$, all $m$ coefficients taken together belong to the
coset $V_l+g$ of $V_l$.  Note that \bean V_l+g & = & V_l + g^{'}
\eean implies \bean \{W_l+g\} \cup \{W_l+\zeta+g\}  & = &
\{W_l+g^{'}\} \cup \{W_l+\zeta+g^{'}\} .  \eean  But this is
impossible since $g, g^{'}$ belong to different cosets of $W_l$
and $g,g^{'}$ have trace zero, whereas, $tr(\zeta)=1$.  It follows
that the coefficient sets of distinct users belong to different
cosets of $V_l$ and are hence distinct.

Thus, the basic sequence $\{ s(g,0,t) \}$ assigned to user $g$ will
take on the form \bean
    s(g,0,t) & = & \sqrt{2\imath} \sum_{k=0}^{m-1}
    2^{m - 1 - k} \imath^{T([1 + 2(g + \delta_k)]\xi^{t + \tau_k})} \ ,
\eean with both $\delta_0$ and $\tau_0$ equal to $0$.

Let $G$ be the set of all such coset representatives of $W_l$ in
$W_{r-1}$. Since each user is associated to a unique coset
representative, the number of users is given by \bean
    \mid G \mid & = & 2^{r-l-1}.
\eean When combined with \eqref{eq:m_1_and_powers_of_2}, we obtain
\bean \frac{2^r}{4(m-1)} \ < \ \mid G \mid \leq  \
\frac{2^r}{2(m-1)}. \eean  Thus the size of $G$ is at most a
factor of $4$ smaller than the best possible suggested by the
Hamming bound, see \eqref{eq:Hamming_GV_bounds}.

Let $\{ \tau_1, \tau_2, \ldots, \tau_{m-1} \}$ be a set of
non-zero, distinct time-shifts with $\{ 1, \alpha^{\tau_1},
\alpha^{\tau_2}, \ldots, \alpha^{\tau_{m-1}} \}$ being a linearly
independent set. Let $\kappa = (\kappa_0, \kappa_1, \ldots,
\kappa_{m - 1}) \in \mathbb{Z}_4 \times \mathbb{F}_2^{m - 1}$.
Family ${\cal SQ}_{M^2}$ is then defined as follows:

\beqa
    {\cal SQ}_{M^2} & = & \left\{
    \left. \left\{ s(g, \kappa, t) \left| \begin{array}{l}
    \kappa \in \mathbb{Z}_4 \times \mathbb{F}_2^{m - 1}
    \end{array} \right\} \right. \right|
    g \in G \right\} \nonumber\\ \label{eqref:Q_M^2_defn}
\eeqa so that each user is identified by an element of $G$.  Each
user is assigned the collection \beqn
    \left\{ s(g, \kappa, t) \ | \ 
    \kappa \in \mathbb{Z}_4 \times \mathbb{F}_2^{m - 1} \right\}
\eeqn of sequences with the $\kappa$-th sequence given by \beqa
    s(g, \kappa, t) & = &
    \sqrt{2 \imath} \left( \sum_{k = 1}^{m - 1}
    2^{m - k - 1} \, \imath^{u_k(t)} (-1)^{\kappa_k} \, +\right. \nonumber\\
    & & \hspace{0.8in} \left.
    2^{m - 1} \, \imath^{u_0(t)}  \right) \imath^{\kappa_0} \label{eqref:mod}
\eeqa where \beqan
    u_0(t) & = & T([1 + 2 g]\xi^t) \nonumber\\
    u_k(t) & = & T([1 + 2 (g + \delta_k)]\xi^{t + \tau_k}), \label{eqref:z4_def}\\
    & & \hspace{1in} k = 1, 2, \ldots, m - 1. \nonumber
\eeqan We will refer to the element $g$ as the {\em ground
coefficient}.  Note that given the ground coefficient and the set
$\{\delta_1, \cdots, \delta_{m-1}\}$, the set of coefficients used
by a user are uniquely determined.  The elements $\{\delta_k\}$
will turn out to provide a selection of the component sequences
that leads to lower correlation values.

\vspace{0.1in}

Within the subset of sequences assigned to a particular user, the
sequences corresponding to $\kappa=0$ will be termed {\em basic
sequences}.   Basic sequences have a simpler representation and
correlations involving basic sequences turn out to be
representative of the general case.

\vspace{0.1in}

\begin{thm} \label{thm:main_pq}
Sequences in Family ${\cal SQ}_{M^2}$ satisfy the following
properties:

\ben

\item \label{thm:main_pq_prd} All sequences in the family have period
$N = 2^r - 1$.

\item \label{thm:main_pq_energy} For large $N$, the energy of any
sequence in the family is given by \beqn
    {\cal E} \approx \frac{2}{3} \, (M^2 - 1) \, N.
\eeqn

\item \label{thm:main_pq_corr} The correlation parameter
$\theta_{\max}$ has the upper bound: \bean
    \theta_{\max} & \lesssim &
    \sqrt{\frac{61}{18}} M^2 \sqrt{N + 1}.
\eeqan For large $M$ and $N$, the normalized maximum correlation
parameter of the family satisfies the bound  \beqn
    \overline{\theta}_{\max} \lesssim   2.76 \sqrt{N}.
\eeqn

\item \label{thm:main_pq_sz} The family can support \bean
\frac{2^r}{4(m-1)} < \ \mid G \mid \ \leq  \frac{2^r}{2(m-1)}
\eean distinct users.  (Note from \eqref{eq:Hamming_GV_bounds}
that this can potentially be improved by a different construction
of the set $G$).

\item \label{thm:main_pq_data_mod}
Each user can transmit $(m + 1)$ bits of data per sequence period.

\item \label{thm:main_pq_ed} The normalized minimum squared Euclidean
distance between all sequences assigned to a user is given by
\beqn
    \overline{d}_{\min}^2 \approx \frac{12}{M^2 - 1} \, N.
\eeqn

\item \label{thm:main_pq_bal} The number ${\bf N}$ of times an element from
the $M^2$-QAM constellation occurs in sequences of large period can
be bounded as: \beqn
    \left| {\bf N} - \frac{N + 1}{M^2} \right| \ \leq \ \frac{M^2 - 1}{M^2} \sqrt{N +
    1} \ ,
\eeqn i.e., the sequences in Family ${\cal SQ}_{M^2}$ are
approximately balanced.

\een
\end{thm}

\vspace{0.1in}

\begin{proof} Property (\ref{thm:main_pq_prd}) follows from the periodicity of the
sequences in Family $\cal A$.  The proof of Property
(\ref{thm:main_pq_energy}) is identical to the proof concerning
the energy of sequences in Family ${\cal CQ}_{M^2}$ (see
Remark~\ref{not:energy}).

Property (\ref{thm:main_pq_corr}) is proved in detail in
Appendix~\ref{app:main_pq_corr_pf}.

Properties (\ref{thm:main_pq_sz}) and (\ref{thm:main_pq_data_mod})
follow directly from the definition of the sequence family.

Property (\ref{thm:main_pq_ed}) can be proved using techniques
similar to those in Appendix~\ref{app:main_cq_ed_pf}; as it turns
out, the minimum Euclidean distance is associated with data sets
$(\kappa,\kappa^{'})$ where \bean \kappa_{m-1} & = &
\kappa_{m-1}^{'} + 2,
\\ \kappa_k & = & \kappa_{k}^{'}, \ \ \ k =0,1,\ldots,(m-2) . \eean

The proof of Property (\ref{thm:main_pq_bal}) concerning symbol
balance is identical to the proof in the case of Family ${\cal
CQ}_{M^2}$.
\end{proof}

\subsection{Variable-Rate Signalling on the Reverse Link Using
Family ${\cal SQ}_{M^2}$} \label{sec:var}

In this section, we show how Families $\{{\cal SQ}_{M^2}\}$ can
also be used to provide variable-rate signalling on a CDMA reverse
link.  We retain the notation of Section~\ref{sec:Fam_PQ_M^2_def}.

We begin by constructing a chain of subspaces \bean
    W_0 \subseteq W_1 \subseteq \cdots \subseteq W_{r-1}
\eean in which each subspace $W_k$ contains only elements of trace
$0$.
Let the elements $\rho_k$ be such that \bean W_{k+1} & = &
W_k \cup \{W_k + \rho_k\} , \eean i.e., $\rho_k$ is a coset
representative of the coset of $W_k$ in $W_{k+1}$ other than $W_k$
itself.

For each $k$, $0 \leq k \leq (r-1)$, the set \bean V_k & = & W_k
\cup (W_k+\zeta) \eean is also a subspace of $\mathbb{F}_q$. Each
element in the coset $W_k+\zeta$ has trace equal to $1$.  Let
$\{\delta_1, \delta_2, \ldots, \delta_{2^{r-1}}\}$  be an ordering
of the elements in $W_{r-1}+\zeta$, obtained by imposing the
condition that the elements of the coset $W_k+\zeta$ precede the
elements of $W_l+\zeta$ if $k < l$.

A user is permitted to pick a c-p $m$ in the range, $1 \leq m \leq
2^{r-1} + 1$, and this choice will permit him to communicate $(m +
1)$ bits per period of the spreading sequence.

Let there be $N_l$ users wishing to communicate using c-p $m$
satisfying \bean \mu(m) & = & l , \eean i.e., associated to sse
$l$.   Our construction below will require that the inequality
\bea
    \sum_{l=0}^{r-1} N_l 2^{l+1} & \leq & 2^r \label{eq:var_sig_count}
\eea hold and we will assume that this is the case.  Let \bean
    l_1 > l_2 > \cdots > l_{K}
\eean be an ordering of subspace-size exponents.    The goal here is
to provide each user with a ground coefficient $g$ which will enable
him to construct his particular QAM sequence.

We begin with the c-ps associated to largest sse $l_1$.  We begin by
partitioning $W_{r-1}$ into disjoint cosets of the subspace
$W_{l_1}$ and assign a coset of $W_{l_1}$ to each of these users.
Each such coset is of the form $g+W_{l_1}$ and the user then
constructs the user's QAM sequence using ground coefficient $g$. The
set of all coefficients assigned to the user, namely the set \bean
\{g, g+\delta_1, \cdots, g+\delta_{m-1} \} \eean then belongs to the
coset of $V_{l_1+1}$ given by \bean V_{l_1+1}+g & = & \{g+W_{l_1}\}
\cup \{(g+\zeta)+W_{l_1}\}. \eean Having in this way made an
assignment of coefficients to the users with largest c-p, we next
move on to the users with next largest c-p. Suppose that
$l_2=l_1-1$. In this case, we can partition one of the unused cosets
of $W_{l_1}$, say $h+W_{l_1}$, according to \bean W_{l_1}+h & = &
\{W_{l_2}+h \} \cup \{ W_{l_2} + \rho_{l_2} +h \} .  \eean We can
then assign either $h$ or $h+\rho_{l_2}$ as the ground coefficient
for the user with sse $l_2$.   If $l_2 =l_1-2$, then we continue the
process by further partitioning each coset $W_{l_1-1}+h$,
$W_{l_1-1}+h+\zeta$ of $W_{l_1-1}$ into two cosets of $W_{l_1-2}$
and assigning a coset of $W_{l_1-2}$ to that user etc. This process
can clearly be continued to satisfy all users provided that the
inequality in \eqref{eq:var_sig_count} is satisfied.

We illustrate with the help of an example for the case $r=4$.

\begin{figure*}[!ht]
    \begin{center}
        \centerline{\epsfig{figure=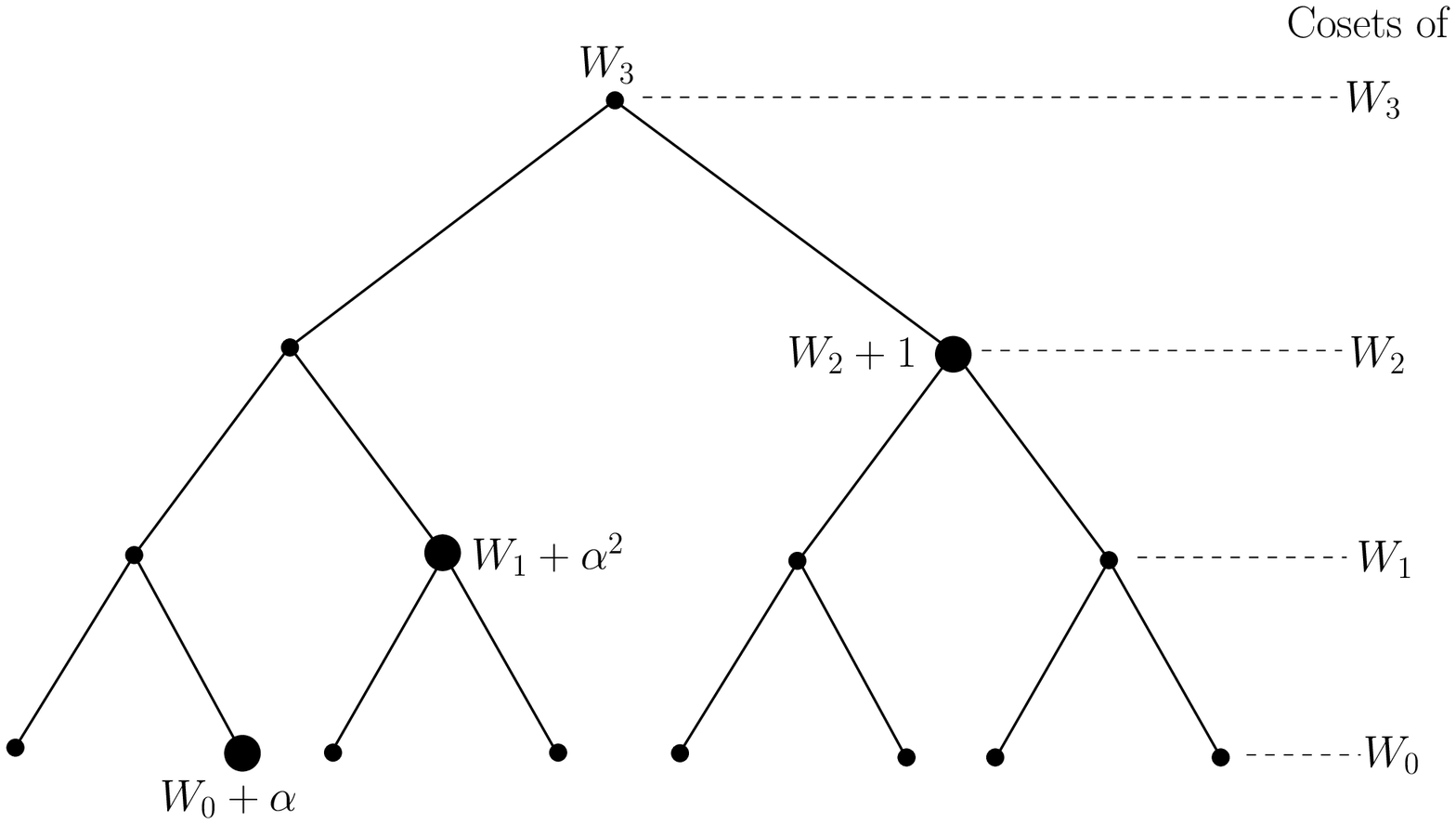,width=120mm}}
        \caption{{Variable-Rate Signalling with Three Users\label{fig:tree_var_rate}}}
    \end{center}
\end{figure*}
\vspace*{0.1in}

\begin{eg}
Let $q=16$ so that $r=4$.   Let the primitive element $\alpha \in
\mathbb{F}_{16}$ satisfy $\alpha^4 + \alpha + 1=0$. It is known that
the element $\alpha^3$ has trace $1$, so we make the selection
$\zeta=\alpha^3$.  Let $W_3, W_3+\zeta$ denote the subsets of
$\mathbb{F}_{16}$ having trace $0$ and $1$ respectively.  Then it
can be verified that \bean
    W_3 & = & \{0,1,\alpha,\alpha^2,\alpha^4,\alpha^5,\alpha^8,\alpha^{10} \} \\
    W_3 + \zeta & = & \{\alpha^3,\alpha^6,\alpha^7,\alpha^9,\alpha^{11},\alpha^{12},\alpha^{13},\alpha^{14}\}.
\eean

Let the values  \beqn
    N_0 \ = \ N_1 \ = \ N_2 \ = \ 1
\eeqn satisfying \eqref{eq:var_sig_count} be given.  The largest
value of $l$ such that $N_l \neq 0$ is $l_1= 2$.  We also have
$l_2=1$, $l_3=0$.  We begin by considering the sequence of
subspaces \bean
    W_0 \ \subseteq \ W_1 \ \subseteq \ W_2.
\eean We choose \bean
    W_0 & = & \{0 \} \\
    W_1 & = & W_0 \cup \{W_0+\alpha\} \ = \ \{0, \alpha \} \\
    W_{2} & = & W_1 \cup \{W_1 + \alpha^2\} \ = \ \{0, \alpha, \alpha^2,\alpha^5 \} .
\eean Thus $\rho_0=\alpha$ and $\rho_1=\alpha^2$. This leads to
\bean
    (\delta_1,\delta_2,\delta_3,\delta_4)
    & = & (\alpha^3,\alpha^9,\alpha^6,\alpha^{11}).
\eean Since $l_1=2$, we begin by considering cosets of $W_2$ in
$W_{r-1}=W_3$. It can be verified that \bean
    W_2 \cup \{W_2+1\} & = & W_{3} .
\eean

We first select the coset $\{W_2+1\}$ (either coset could have been
chosen at this step).  The corresponding ground coefficient equals
$1$ and this is assigned to the user with sse$ \ = \ l_1 \ = \ 2$.
Since there is only one user with sse equal to $2$, we move on to
consider the user with sse $= \ l_2 \ =$ 1.  Our next step is to
partition the remaining coset of $W_2$, namely, in this case, $W_2$
itself. Since $\rho_1=\alpha^2$, we can partition $W_2$ into \bean
W_2 & = & W_1 \cup \{ W_1+\alpha^2\} . \eean

Again faced with a choice, we choose to assign coset
$W_1+\alpha^2$ to the user with sse $l_2=1$, corresponding to
choice of $\alpha^2$ as the ground coefficient. This leaves us
with the coset $W_1$ .  There is one remaining user with sse
$=l_0=1$. Since $\rho_0=\alpha$, we have the partitioning \bean
W_1 & = & W_0 \cup \{W_0+\alpha\} .  \eean

Again we choose to assign coset $W_0+\alpha$ to the last remaining
user, whose ground coefficient thus is set equal to $\alpha$.

Thus the signals of the $3$ users are given by \bean
    s(1,0,t) & = & \sqrt{2\imath} \left( \imath^{T([1+2 \alpha^{12}]x \xi^{\tau_4})} +
    2 \imath^{T([1+2\alpha^{13}]x \xi^{\tau_3})} + \right. \\
    & & \left. + 4 \imath^{T([1+2 \alpha^{7}]x \xi^{\tau_2})} +
    8 \imath^{T([1+2\alpha^{14}]x \xi^{\tau_1})}\right.  \\
    & & \left.+ 16\imath^{T([1+2]x)} \right), \\
    s(\alpha^2,0,t)  & = & \sqrt{2\imath} \left(
    \imath^{T([1+2\alpha^{11}] x\xi^{\tau_2})} + 2
    \imath^{T([1+2\alpha^{6}]x \xi^{\tau_1})} + \right. \\
    & & \left. + 4 \imath^{T([1+2 \alpha^{2}]x)} \right) , \\
    s(\alpha,0,t)  & = & \sqrt{2\imath} \left( \imath^{T([1+2\alpha^9]x \xi^{\tau_1}})
    + 2 \imath^{T([1+2\alpha]x)} \right),
\eean where $x=\xi^t$ and $(\tau_1, \tau_2, \tau_3, \tau_4) = (1, 2,
3, 4)$.

Figure~\ref{fig:tree_var_rate} graphically depicts the assignment
of ground coefficients.  In the tree, the root node corresponds to
the subspace $W_{r-1} = W_3$ in the example.  Each node in the
tree corresponds to a coset $W_l+g$ of some subspace $W_l$ of
$W_3$.  The nodes one level down from the root node corresponds to
the two cosets of $W_2$ (one of them of course is $W_2$ itself).
The nodes two levels down from the root node correspond to cosets
of $W_1$. The leaf nodes correspond to cosets $W_0+g$ of
$W_0=\{0\}$ in $W_3$. Each user is assigned a distinct node in the
tree.  The ground coefficient assigned to the particular user can
be chosen to be any coset representative of the coset associated
to that node. Given that a node is assigned to a user, no
descendant of that node can be assigned to any other user.  The
coefficients assigned to the user are of the form \bean \{g,
g+\delta_1, \cdots, g+\delta_{m-1}\}. \eean  The tree only depicts
how $g$ is to be selected.  Given $g$, the remaining coefficients
are obtained by adding elements $\delta_k$ to $g$.  The elements
$\{ \delta_k\}$ are themselves drawn from the coset $W_3+\zeta$ of
$W_3$. This coset is not depicted in the tree.

In this example, the reader will have noticed that there are
$16-10=6$ unused sequences remaining in Family ${\cal A}$. These may
be added to the existing list of sequences as users wishing to use a
$4$-QAM constellation: \bean s(0,0,t) & = & \sqrt{2\imath} \left(
\imath^{T(x)}  \right) \\
s(\alpha^{3},0,t) & = & \sqrt{2\imath} \left(
\imath^{T([1+2\alpha^{3}]x)} \right) \\
 s(\alpha^4,0,t) & = &
\sqrt{2\imath} \left(
\imath^{T([1+2\alpha^4]x)}  \right) \\
s(\alpha^5,0,t) & = & \sqrt{2\imath} \left(
\imath^{T([1+2\alpha^5]x)}  \right) \\
s(\alpha^8,0,t) & = & \sqrt{2\imath} \left(
\imath^{T([1+2\alpha^8]x)}  \right) \\
s(\alpha^{10},0,t) & = & \sqrt{2\imath} \left(
\imath^{T([1+2\alpha^{10}]x)}  \right). \eean

\end{eg}

\section{An Interleaved Construction for $16$-QAM} \label{sec:qam_16}

Setting $M=4$ in the ${\cal SQ}_{M^2}$ construction yields Family
${\cal SQ}_{16}$.   The associated normalized maximum correlation
parameter $\overline{\theta}_{\max}$ for this family is upper
bounded by $1.61 \sqrt{N}$ which is lower than the upper bound on
$\overline{\theta}_{\max}$ of $1.8 \sqrt{N}$ for sequence family
${\cal Q}_{B}$. In the next subsection, we interleave sequences to
construct a $16$-QAM sequence family whose upper bound on
$\overline{\theta}_{\max}$ is further lowered to $\sqrt{2N}$.

\subsection{Family ${\cal IQ}_{16}$} \label{sec:fam_qi_16}

Let $\delta_1$ be an element of $\mathbb{F}_q$ with $tr(\delta_1)=1$
and $H$ the additive subgroup $\{0,\delta_1\}$.
 Let $G$ be the set obtained by picking one coset representative
 from each of the coset representatives of $H$ in $\mathbb{F}_q$.
 Thus $G$ is of size
 \bean
\mid G \mid & = & \frac{q}{2}.
 \eean
Let $\tau_1$ be such that $\{1,\alpha^{\tau_1}\}$ is a linearly
independent set over $\mathbb{F}_2$.

Family ${\cal IQ}_{16}$ is then defined as the collection of
sequences: \beq
    {\cal IQ}_{16} = \{
    \{ s(g, \kappa, t) \ | \ \kappa \in {\mathbb{Z}}_4 \times
    {\mathbb{F}}_2 \} \ | \ g \in G \} \label{eqref:qam_fam_a}
\eeq with the $\kappa$-th sequence assigned to the $g$-th user sequence given
by \eqref{eq:seq_fam_a}.
\begin{figure*}
\beq
    s(g, \kappa, t) = \left\{
    \begin{array}{ccc}
    \sqrt{2 \imath} \left( \imath^{u_1(t)} (-1)^{\kappa_1} \, +
    2 \, \imath^{u_0(t)}  \right) \imath^{\kappa_0}, &  & t \ \mbox{even} \\
    \sqrt{2 \imath}  \, \imath \left( \imath^{u_0(t)} \, -
    2 \, \imath^{u_1(t) } (-1)^{\kappa_1}  \right)
    \imath^{\kappa_0}, &  & t \ \mbox{odd.}
    \end{array} \right.  \label{eq:seq_fam_a}
\eeq
\end{figure*}
Sequences $\{u_0(t)\}, \{u_1(t)\}$ are given by \beqan
    u_0(t) & = & T([1 + 2 g]\xi^t) \label{eq:u_0} \\
    u_1(t) & = & T([1 + 2 (g + \delta_1)]\xi^{t + \tau_1}). \label{eq:u_1}
\eeqan

The theorem below identifies the principal properties of Family
${\cal IQ}_{16}$.

\vspace{0.1in}

\bthm \label{thm:main_iq_16} Let ${\cal IQ}_{16}$ be the family of
sequences over $16$-QAM constellation defined in
(\ref{eqref:qam_fam_a}). Then,

\ben

\item \label{thm:main_iq_16_prd} All sequences in ${\cal IQ}_{16}$ have period $N = 2(2^r - 1)$.

\item \label{thm:main_iq_16_energy}
For large values of $N$, the energy of the sequences in the family is given by
\beqn
    {\cal E} \approx 10 \, N.
\eeqn

\item \label{thm:main_iq_16_corr}
For large values of $N$, the normalized maximum correlation parameter of the
family can be bounded as \beqn
    \overline{\theta}_{\max} \lesssim \sqrt{2} \sqrt{N}.
\eeqn

\item \label{thm:main_iq_16_size}
Family ${\cal IQ}_{16}$ can support $(N + 2)/4$ distinct users.

\item \label{thm:main_iq_16_data_mod}
Each user can transmit $3$ bits of data per sequence period.

\item \label{thm:main_iq_16_ed}
The normalized minimum squared Euclidean distance between all sequences
assigned to a user is given by \beqn
    \overline{d}_{\min}^2 \approx 2 N.
\eeqn

\item \label{thm:main_iq_16_bal}
The sequences in Family ${\cal IQ}_{16}$ are approximately balanced.

\een \ethm

\vspace*{0.2in}

\begin{proof} A proof of the Property (\ref{thm:main_iq_16_corr}) can be found in
Appendix~\ref{app:main_iq_16_corr_pf}. The remaining properties
can be established in essentially the same manner as was done in
the case of Family ${\cal SQ}_{M^2}$.
\end{proof}


\section{Families of sequences over Q-PAM constellation} \label{sec:am_psk}

So far all our constructions have been for sequences over the
$M^2$-QAM constellation.  In this section, we shall show that by
restricting the symbol alphabet to a size $2M$ subset of the QAM
constellation, the maximum correlation magnitude can be further
lowered.  This subset of the $M^2$-QAM constellation is given by
\eqref{eq:AM_PSK_constellation} and for obvious reasons, will be
referred to as the quadrature-PAM or Q-PAM constellation.

We first present a general technique for constructing families
${\cal P}_{2M}$ of sequences over the Q-PAM constellation.
Subsequently, we shall use a different interleaving technique to
construct a family ${\cal IP}_8$ of sequences over the specific
$8$-ary Q-PAM constellation.   Remarkably, this latter sequence
family achieves the Welch bound \cite{Wel} on maximum magnitude of
correlation with equality.  To our knowledge, this family is the
only-known non-trivial optimal family of sequences over a non-PSK
symbol alphabet, i.e., over an alphabet not comprised of roots of
unity.

\subsection{Family ${\cal P}_{2M}$}
\label{sec:q_pam}

Let $\delta_1, \delta_2, \ldots, \delta_{m - 1}$ be trace $1$
elements of $\mathbb{F}_q$ having the property that $\{1,
\delta_1, \delta_2, \ldots, \delta_{m - 1} \}$ is a linearly
independent set over $\mathbb{F}_2$.  Set $\delta_0=0$.
Let $G =\{g_a\}$ be the largest subset of $\mathbb{F}_q$ having
the property that \bean g_a + \delta_k & \neq & g_b+\delta_l, \ \
\ g_a, g_b \in G, \ \ 0 \leq k,l \leq (m-1), \eean unless $g_a =
g_b$ and $\delta_k = \delta_l$. As before, the corresponding
Gilbert-Varshamov and Hamming bounds on the size of $G$ are given
by \bea
    \frac{2^{r}}{1+\binom{m-1}{1}+\binom{m-1}{2}}\leq & \mid G
    \mid & \leq \frac{2^r}{{1+\binom{m-1}{1}}} \ .   \label{eq:PAM_GV-Hamming}
\eea

Family ${\cal P}_{2M}$ is defined as follows: \beqa
    {\cal P}_{2M} & = & \left\{
    \left. \left\{ s(g, \kappa, t) \left|
    \kappa \in {\mathbb{Z}}_4 \times {\mathbb{F}}_2^{m - 1}
    \right\} \right. \right|
    g \in G \right\}. \nonumber\\ \label{eqref:am_psk_M^2_defn}
\eeqa Each user is thus assigned the set \beqn
    \left\{ s(g, \kappa, t) \ | \
    \kappa \in {\mathbb{Z}}_4 \times {\mathbb{F}}_2^{m - 1} \right\}
\eeqn of sequences with the $(\kappa)$-th sequence given by \beqan
    s(g, \kappa, t) & = &
    \sqrt{2 \imath} \left( \sum_{k = 1}^{m - 1}
    2^{m - k - 1} \, \imath^{u_k(t)} (-1)^{\kappa_k} \, +\right. \nonumber\\
    & & \hspace{0.8in} \left.
    2^{m - 1} \, \imath^{u_0(t)}  \right) \imath^{\kappa_0}
\eeqan where \beqan
    u_0(t) & = & T([1 + 2 g]\xi^t) \nonumber\\
    u_k(t) & = & T([1 + 2 (g + \delta_k)]\xi^t), \\
    & & \hspace{1in} k = 1, 2, \ldots, m - 1. \nonumber
\eeqan

Note that in relation to the definition of the sequence Family
${\cal SQ}_{M^2}$, the time-shift parameters $\tau_k$ are absent in
the present construction.  It is the absence of the terms
$\xi^{\tau_k}$ as we inadvertently discovered, that causes the
sequence symbol alphabet to lie in the Q-PAM subconstellation.
Nevertheless, as we see below, Family ${\cal P}_{2m}$ has
essentially the same properties as does Family ${\cal SQ}_{M^2}$
while enjoying the added advantage of a lower value of maximum
correlation magnitude.  The principal properties of Family ${\cal
P}_{2M}$ are summarized in the following theorem:

\vspace{0.1in}

\bthm \label{thm:main_am_psk} Let $m \geq 2$ be a positive integer
and let ${\cal P}_{2M}$ be the family of sequences over $2M$-ary
Q-PAM constellation defined in (\ref{eqref:am_psk_M^2_defn}). Then,

\ben

\item \label{thm:main_am_psk_prd} All sequences in ${\cal P}_{2M}$ have period $N =
2^r - 1$.

\item \label{thm:main_am_psk_energy} For large values of $N$, the energy of the
sequences in the family is given by \beqn
    {\cal E} \approx \frac{2}{3} \, (M^2 - 1) \, N.
\eeqn

\item \label{thm:main_am_psk_corr}  The maximum correlation parameter of the family
can be bounded as \beqan
    \theta_{\max} \ \lesssim \ \sqrt{\frac{20}{9}} M^2 \sqrt{N + 1}.
\eeqan

For large values of $M$ and $N$, the normalized maximum correlation
parameter of the family can be bounded as \beqn
    \overline{\theta}_{\max} \lesssim \sqrt{5} \, \sqrt{N}.
\eeqn

\item \label{thm:main_am_psk_sz_mod} Family ${\cal P}_{2M}$ can
support $|G|$ distinct users where $| G |$ lies in the range given
in \eqref{eq:PAM_GV-Hamming}.

\item Each user can transmit $m + 1$ bits of data per sequence
period.

\item \label{thm:main_am_psk_ed} The normalized minimum squared Euclidean distance
between all sequences assigned to a user is given by \beqn
    \overline{d}_{\min}^2 \approx \frac{12}{M^2 - 1} \, N.
\eeqn

\item The sequences in Family ${\cal P}_{2M}$ are approximately balanced.

\een \ethm

\vspace{0.1in}

\bpf The above properties of Family ${\cal P}_{2M}$ can be
established using the same techniques used to prove properties of
Families ${\cal CQ}_{M^2}$ and ${\cal SQ}_{M^2}$, and are hence
omitted.

The only difference in the correlation computations for Families
${\cal P}_{2M}$ and ${\cal SQ}_{M^2}$ is that, in this case,
$\theta_{u_k,u_0^{'}}(\tau)$ is also at right angles with
$\theta_{u_0,u_k^{'}}(\tau)$.  This is in addition to
$\theta_{u_0,u_0^{'}}(\tau)$ being in right angles with
$\theta_{u_k,u_k^{'}}(\tau)$ (see
Appendix~\ref{app:main_pq_corr_pf}).\epf

\vspace*{0.1in}

\subsubsection{Variable-Rate Signalling with Family ${\cal P}_{2M}$}

In Section~\ref{sec:var}, we described in detail a technique to
allow different users to transmit at variable rates by choosing
sequences from various members of Families $\{ {\cal SQ}_{M^2} \}$
corresponding to constellations of different sizes. Similar
techniques can be used to permit variable-rate signalling on the
reverse link of a CDMA system in which users are permitted to choose
spreading sequences from Families ${\cal P}_{2M}$ for different
values of parameter $M$. One obvious difference from the previous
case is the linear independence of the set $\{ 1, \delta_1, \ldots,
\delta_{m - 1}\}$. A key ingredient of the sequence assignment in
the variable-rate signalling scheme involving Family ${\cal
SQ}_{M^2}$ was the identification, for every collection $H$ of
elements $H=\{\delta_1, \delta_2,\cdots \delta_{m-1}\}$, of the
smallest subspace $V_{l+1}$ containing $H$. In the case of Family
${\cal SQ}_{M^2}$, this subspace was of dimension $l+1$ where
$l=\mu(m)$. In the present case, the linear independence of the
$\{\delta_j\}$ forces $l+1=(m-1)$ and the choice \bean V_{l+1} & = &
\left<\delta_1, \delta_2, \cdots, \delta_{m-1} \right> .  \eean
 Given the subspaces $V_{l+1}$ the assignment proceeds as earlier.
  We omit the details.

\begin{figure*}[!ht]
\beq
    s(g, \kappa, t) = \left\{
    \begin{array}{ccc}
    \sqrt{2 \imath} \left( \imath^{u_1(t)} (-1)^{\kappa_1} \, +
    2 \, \imath^{u_0(t)}  \right) \imath^{\kappa_0} & , & t \ \mbox{even} \\
    \sqrt{2 \imath}  \, \imath \left( \imath^{u_0(t)}  \, -
    2 \, \imath^{u_1(t)}  (-1)^{\kappa_1}  \right)
    \imath^{\kappa_0} & , & t \ \mbox{odd}
    \end{array} \right. \label{eq:eq_opt}
\eeq \end{figure*}

\begin{table*}[!ht]
    \caption{Simulation results for various sequence families.}
    \label{tab:sim}
\begin{center}
\begin{tabular}{|c|c|c|c|c|c|c|c|c|}
    \hline \hline
    Family & Constellation  & Period  & Family Size  & Data
    & ${\theta}_{max}$ & $\frac{\overline{\theta}_{max}}{\sqrt{N}}$
    & ${d}_{\min}^2$ & $\frac{\overline{d}_{\min}^2}{N}$ \\
    & & $(N)$ & Rate & & & & & \\
    \hline
    ${\cal IQ}_{16}$ & $16$-QAM & $30$ & $8$ & $3$ & $100$ & $1.82$ & $600$ & $2$ \\
    \hline
    ${\cal SQ}_{16}$ & $16$-QAM & $15$ & $8$ & $3$ & $82.38$ & $2.04$ & $120$ & $0.75$ \\
    \hline
    ${\cal IP}_{8}$ & $8$-ary Q-PAM & $30$ & $8$ & $3$ & $72.11$ & $1.31$ & $600$ & $2$ \\
    \hline
    ${\cal P}_{8}$ & $8$-ary Q-PAM & $15$ & $8$ & $3$ & $66.48$ & $1.81$ & $120$ & $0.79$\\
    \hline
    ${\cal CQ}_{16}$ & $16$-QAM & $15$ & $8$ & $4$ & $84.21$ & $2.1$ & $60$ & $0.34$\\
    \hline \hline
\end{tabular}
\end{center}
\end{table*}

\subsection{Family ${\cal IP}_{8}$}

There are not many families of sequences that (asymptotically)
meet the Welch lower bound \cite{Wel} on sequence correlation, see
\cite{HelKum}.   To the authors' knowledge, those that do achieve
the Welch bound with equality, are over a signal constellation
associated to $M$-ary phase-shift keying for some $M \geq 2$.  The
asymptotically optimal family of sequences constructed in this
section, Family ${\cal IP}_{8}$, is, however, over the $8$-ary
Q-PAM alphabet (see Fig.~\ref{fig:am_psk}) and is constructed
using sequence interleaving.

Let $\delta_1$ be a trace $1$ element of $\mathbb{F}_q$ such that
$\{1, \delta_1\}$ is a linearly independent set over
$\mathbb{F}_2$. Let $H \ = \ \{0, \delta_1\}$ denote the additive
subgroup generated by $\delta$ and let $G$ be the set of coset
representatives of $H$ in $\mathbb{F}_q$.

Family ${\cal IP}_{8}$ is defined as follows: \beq
    {\cal IP}_{8} \ = \ \{
    \{ s(g, \kappa, t) \ | \ \kappa \in {\mathbb{Z}}_4 \times {\mathbb{F}}_2 \} \ | \ g \in G \}
    \label{eqref:fam_opt}
\eeq with the $\kappa$-th sequence given by \eqref{eq:eq_opt} and
with \beqan
    u_0(t) & = & T([1 + 2 g)]\xi^t) \\
    u_1(t) & = & T([1 + 2 (g + \delta_1)]\xi^t).
\eeqan

As would be evident from the definition of the sequence family, we
have interleaved two sequences over the $8$-ary Q-PAM alphabet to
generate a single sequence over the same alphabet.

Most of the properties of Family ${\cal IP}_{8}$ like period of
the sequences, family size, data rate, Euclidean distance, and
balance are exactly the same as that of Family ${\cal IQ}_{16}$
(see Section~\ref{sec:fam_qi_16}).  The main difference between
the two families is that in case of Family ${\cal IP}_{8}$, for
large values of $N$, the normalized maximum correlation parameter
achieves the Welch bound, i.e., can be bounded as \beqn
    \overline{\theta}_{\max} \lesssim \sqrt{N} \ ,
\eeqn and this is established in Appendix~\ref{app:welch_bd_pf}.
For sake of brevity, we omit the proofs of the other properties.

\section{Simulation Results}

We have simulated one member of each of the various sequence
families that we have constructed in this paper. The results of the
simulation are available in Table~\ref{tab:sim}. The underlying
finite field that is used to construct all the families is the same:
${\mathbb F}_{16}$.  We have chosen the minimal polynomial $X^4 + X
+ 1$ to construct ${\mathbb F}_{16}$. The construction of the
families mirrors their definition in the paper. Since we have chosen
sequences of short period, the results are not completely indicative
of the asymptotic behavior of the families.

\appendices

\section{Closed-Form Expression for Family ${\cal A}$ Correlation }
\label{app:closed_form_Fam_A}

A useful closed-form expression for the pairwise-correlation between
a pair of sequences drawn from Family ${\cal A}$ is given below.

\vspace{0.1in}

\blem \label{lem:Gamma} \cite{YanHelKumSha} Let $a + 2 b \in R$
with $a, b \in \cal T$, $a \neq 0$. Define \beqn
    \Gamma(a + 2 b) := \sum_{x \in \cal T} \imath^{T([a + 2 b]x)}.
\eeqn Then \beqa \Gamma(1) & = &
    \left\{ \begin{array}{ccc}
        \sqrt{2^r} \ \epsilon^r ,& & r \ \mbox{is odd} \\
        -\sqrt{2^r} \ \epsilon^r ,& & r \ \mbox{is even} \\
    \end{array}
    \right. \label{eqref:gamma_1}
    \eeqa
where \beqn
    \epsilon \ = \ \frac{1 + \imath}{\sqrt{2}}.
\eeqn  Further,
    \beqa
    \Gamma(a + 2 b) & = & \Gamma(1) \, \imath^{-T(\frac{b}{a})}.
    \label{eqref:gen_Gamma}
\eeqa

\vspace{0.1in}

\bpf A proof of \eqref{eqref:gen_Gamma} can be found in
\cite{YanHelKumSha} and is presented below for the sake of
completeness. \beqa
    \Gamma(a + 2 b) & = &  \sum_{x \in \cal T} \imath^{T([a + 2 b]x)} \nonumber\\
    & = & \sum_{x \in \cal T} \imath^{T([1 + 2 \frac{b}{a}]a x)} \nonumber\\
    & = & \sum_{x \in \cal T} \imath^{T([1 + 2 \frac{b}{a}]x)} \nonumber\\
    & = & \Gamma(1 + 2 \gamma), \ \text{ where } \gamma = \frac{b}{a} \in \cal T.
    \label{eqref:Gamma_arb}
\eeqa The third equation is obtained by replacing $x$ by $a x$. Now,
if $x$ and $\gamma$ are two elements in $\cal T$, then $(x + \gamma
+ 2 \sqrt{x \gamma})$ also belongs to $\cal T$. If $x$ runs over all
elements of $\cal T$, then $(x + \gamma + 2 \sqrt{x \gamma})$ also
runs over all elements of $\cal T$. Therefore \beqa
    \Gamma(1) & = & \sum_{x \in \cal T} \imath^{T(x)} \nonumber\\
    & = & \sum_{x \in \cal T} \imath^{T(x + \gamma + 2 \sqrt{x \gamma})}
    \nonumber\\
    & = & \sum_{x \in \cal T} \imath^{T(x + 2 \sqrt{x \gamma})} \imath^{T(\gamma)}
    \nonumber\\
    & = & \sum_{x \in \cal T} \imath^{T(x + 2 x \gamma)} \imath^{T(\gamma)}
    \nonumber\\
    & = & \sum_{x \in \cal T} \imath^{T([1 + 2 \gamma]x)} \imath^{T(\gamma)}
    \nonumber\\
    & = & \Gamma(1 + 2 \gamma) \imath^{T(\gamma)}.
    \label{eqref:Gamma_1_chk}
\eeqa Proof of the lemma is completed by comparing the two
expressions in (\ref{eqref:Gamma_arb}) and
(\ref{eqref:Gamma_1_chk}).

We refer the reader to \cite{YanHelKumSha} for the
algebraic-geometric proof of \eqref{eqref:gamma_1}.   \epf \elem

\vspace{0.1in}

The following lemma summarizes the correlation properties of
sequences from Family $\cal A$.

\vspace{0.1in}

\blem \label{lem:Fam_A} Consider two sequences from Family $\cal
A$ defined as \beqan
    s(a, t) & = & \imath^{T([1 + 2 a] \xi^t)} \ \ \ \text{and} \\
    s(b, t) & = & \imath^{T([1 + 2 b] \xi^t)}, \  \ a, b \in \cal T.
\eeqan Then \beqan
    \theta_{s(a), s(b)}(\tau) & = & \left\{ \begin{array}{ll}
    2^r - 1, & a = b, \, \tau = 0 \\
    -1, & a \neq b, \, \tau = 0 \\
    -1 + \Gamma(1) \imath^{-T(z)}, & \tau \neq 0
    \end{array} \right.
\eeqan where \beqn
    z = a + \frac{a + b}{y} + \frac{1}{\sqrt{y}} +
            2 \mu(a,b,y)
\eeqn with \beqn
    y = \xi^{\tau} + 1 + 2 \sqrt{\xi^{\tau}}  \nonumber
\eeqn and $\mu$ some function of $a,b,y$.

\bpf Set $x=\xi^{\tau}$.  The proof follows from an application of
Lemma~\ref{lem:Gamma}, cf. \cite{BozHamKum,YanHelKumSha}, and noting
that \bean
    \lefteqn{[1+2a]x -[1+2b]}\\
    & = & (x-1)+2(ax+b) \\
    & = & x+1+2 \sqrt{x} + 2[ax+b+1+\sqrt{x}] \\
    & = & y+2[a(y+1)+b+1+\sqrt{y}+1] \\& = & y+ 2[ay+a+b+\sqrt{y}] \ ,
\eean where we have set \[ y \ \triangleq \ x+1+2 \sqrt{x}. \]

Thus the ``$\gamma$'' in Lemma~\ref{lem:Gamma}, is given by \beqn
    a+\frac{a+b}{y}+\frac{1}{\sqrt{y}}+2\mu(a,b,y).
\eeqn \epf \elem

\section{Balance of Sequences over $M^2$-QAM}
\label{app:bal_gen_pf}

\noindent (Proof of Lemma~\ref{lem:bal_gen}).

\vspace*{0.2in}

Let ${\bf N}(\nu)$ be the number of times the element $\sum_{k =
0}^{m - 1} 2^k \imath^{\nu_k}$ from the $M^2$-QAM constellation
occurs in one period of the sequence $\{ s_p(t) \}$, where $\nu =
(\nu_0, \nu_1, \ldots, \nu_{m - 1})$.  We have \bean
    {\bf N}(\nu) & = & \left| \left\{ x \in {\cal T} \ | \
    T(a_{p, k}x) = \nu_k , \ k = 0, 1, \ldots, m - 1
    \right\} \right|
\eean We rewrite the expression for ${\bf N}(\nu)$ with the aid of
exponential sums to get \beqa
    \lefteqn{{\bf N}(\nu)} \nonumber\\
    & = & \frac{1}{4^m} \, \sum_{x \in \cal T}
    \sum_{\omega \in \mathbb{Z}_4^m}
    \imath^{\sum_{k = 0}^{m - 1} (T(a_{p, k} x) - \nu_k)\omega_k},
    \nonumber\\
    & = & \frac{1}{M^2} \, \sum_{\omega \in \mathbb{Z}_4^m}
    \imath^{- \sum_{k = 0}^{m - 1} \nu_k \omega_k} \cdot  \nonumber\\
    & & \hspace{0.5in}  \sum_{x \in \cal T}
    \imath^{ T(\left[ \sum_{k = 0}^{m-1} \omega_k a_{p, k} \right] x)} \nonumber\\
    & = & \frac{q}{M^2} +
    \frac{1}{M^2} \sum_{\omega \in \mathbb{Z}_4^m, \ \omega \neq 0}
    \imath^{- \sum_{k = 0}^{m - 1} \nu_k \omega_k} \cdot \nonumber\\
    & & \left(\sum_{x \in \cal T}
    \imath^{ T \left( \left[ \sum_{k=0}^{m-1} \omega_k a_{p,k} \right] x \right)}
    \right) . \label{eqref:count_M^2}
\eeqa By the linear independence of the coefficients $a_{p, k}$,
$\sum_{k = 0}^{m - 1} \omega_k a_{p,k}$ does not equal $0$ for any
choice of $\omega$ and, hence, the magnitude of \bean
    \left(\sum_{x \in \cal T}
    \imath^{ T \left( \left[ \sum_{k=0}^{m-1} \omega_k a_{p,k} \right] x \right)}
    \right)
\eean is bounded from above by $\sqrt{q} \ = \ \sqrt{N+1}$ (see
Appendix~\ref{app:closed_form_Fam_A}). With this, we get the bound
\bean
    \left| {\bf N}(\nu) \ - \  \frac{q}{M^2} \right|
    & \leq &  \frac{M^2 - 1}{M^2} \sqrt{N+1},
\eean  thus proving approximate balance.

\section{Energy of Sequences in Canonical Family ${\cal
CQ}_{M^2}$} \label{app:main_cq_energy_pf}

\noindent (Proof of Property (\ref{thm:main_cq_energy}) in
Theorem~\ref{thm:main_cq})

\vspace*{0.2in}

Consider two sequences in Family ${\cal CQ}_{M^2}$, $\{s(g, \kappa,
t)\}$ and $\{s(g^{'}, \kappa^{'}, t)\}$, given by \bea
    s(g, \kappa, t) & = & \sqrt{2 \imath} \sum_{k = 0}^{m - 1} 2^k
    \imath^{u_k(t)} \imath^{\kappa_k} \label{eq:s_g_cq}\\
    s(g^{'}, \kappa^{'}, t) & = & \sqrt{2 \imath} \sum_{l = 0}^{m - 1} 2^l
    \imath^{u_l^{'}(t)} \imath^{\kappa_l^{'}} . \label{eq:s_g'_cq}
\eea  The correlation between the two sequences is given by \bean
    \lefteqn{ \theta_{s(g,\kappa),s(g',\kappa')}(\tau) =} \\
    & & 2 \sum_{k = 0}^{m - 1} \sum_{l = 0}^{m - 1} 2^{k+l}
    \imath^{u_k(t+\tau) - u_l^{'}(t)} \imath^{\kappa_k-\kappa^{'}_l} .
\eean

Thus, energy of a sequence $\{ s(g, \kappa, t) \}$ is given by \bean
    \lefteqn{ {\cal E}(s(g,\kappa)) } \\
    & = & \theta_{s(g,\kappa),s(g,\kappa)}(0) \\
    & = & 2 \sum_t \sum_k \sum_l 2^{k+l} \imath^{u_k(t)-u_l(t)}
    \imath^{\kappa_k-\kappa_l}   \\
    & = & 2 \sum_{k=0}^{m-1}4^k \, N + 2 \sum_{k,l,k \neq l}
    2^{k+l} \theta_{u_k, u_l}(0) \imath^{\kappa_k-\kappa_l} .
\eean

It follows from this that \bean
    \lefteqn{ \left| {\cal E}(s(g,\kappa)) - 2 \frac{(M^2-1)}{3}N \right| } \\
    & \leq & 2 \left( \sum_{k,l}2^{k+l} - \sum_{k=l}2^k \right) \sqrt{N+1} \\
    & = & 2 \left( (2^m - 1)^2 - \frac{4^m - 1}{3} \right) \sqrt{N+1} \\
    & = & 2 \left( \frac{2}{3} 4^m - 2^{m+1} + \frac{4}{3} \right\}
    \sqrt{N+1}.
\eean

Thus, for large $N$, the energy is approximately given by \bean
    {\cal E}(s(g,\kappa)) & \approx & 2 \frac{(M^2-1)}{3}N.
\eean

\section{Correlation of Sequences in Canonical Family ${\cal
CQ}_{M^2}$} \label{app:main_cq_corr_pf}

\noindent (Proof of Property (\ref{thm:main_cq_corr}) in
Theorem~\ref{thm:main_cq})

The correlation between two sequences, $\{s(g,\kappa, t)\}$ and
$\{s(g^{'}, \kappa^{'}, t)\}$ (see \eqref{eq:s_g_cq} and
\eqref{eq:s_g'_cq}), from Family ${\cal CQ}_{M^2}$ is given by \bean
    \lefteqn{ \theta_{s(g,\kappa),s(g',\kappa')}(\tau) = } \\
    & & 2 \sum_t \sum_k \sum_l 2^{k+l} \imath^{u_k(t+\tau)-u_l^{'}(t)}
    \imath^{\kappa_k-\kappa^{'}_l} .
\eean

Thus, \bean
    \lefteqn{ \left| \theta_{s(g,\kappa),s(g^{'},\kappa^{'})}(\tau) \right| }\\
    & \leq & 2 \sum_k \sum_l 2^{k+l}
    \left| \theta_{u_k, u_l^{'}}(\tau) \right| \\
    & \leq & 2 \sum_{k,l} 2^{k+l} (1 + \sqrt{N+1}) \\
    & \lesssim & 2 (2^m - 1)^2 \sqrt{N+1} \\
    & = & 2(M - 1)^2 \sqrt{N+1} .
\eean

Normalizing with the energy of the concerned sequences, we obtain
\bean
    \overline{\theta}_{\max}
    & \lesssim & \frac{2 (M-1)^2 }{2 (M^2-1)/3} \sqrt{N+1} \\
    & \lesssim & \frac{3(M-1)^2}{(M^2-1)} \sqrt{N+1} \\
    & \lesssim & 3 \sqrt{N}
\eean for large $N$ and $M$.

\section{Minimum-Squared Euclidean Distance for Canonical Family ${\cal
CQ}_{M^2}$} \label{app:main_cq_ed_pf}

\noindent (Proof of Property (\ref{thm:main_cq_ed}) in
Theorem~\ref{thm:main_cq})

We are interested in computing the minimum squared Euclidean
distance between all the sequences assigned to the same user.
Consider the two sequences assigned to a user: $\{ s(g,\kappa, t)
\}$ and $\{ s(g,\kappa^{'}, t) \}$.

Let $L$ be the set of all indices $k$ such that \beqn
    \kappa_k \ \neq \ \kappa_k^{'} .
\eeqn  Then \bean
    \lefteqn{ d_E^2(s(g,\kappa, t), s(g,\kappa^{'}, t)) }\\
    & = & 2 \sum_t \left| \sum_{k \in L}  2^k \imath^{u_k(t)}
    (\imath^{\kappa_k}-\imath^{\kappa_k^{'}}) \right|^2 \\
    & = & 2 \sum_t \sum_{k,l \in L} 2^{k+l}
    \imath^{u_k(t)- u_l(t)} \cdot \\
    & & \hspace{0.6in}
    (\imath^{\kappa_k}-\imath^{\kappa_{k}^{'}})
    (\imath^{-\kappa_l}-\imath^{-\kappa_{l}^{'}})\\
    & = & 2 \sum_t \sum_{k \in L} 4^k
    \left| \imath^{\kappa_k}-\imath^{\kappa_{k}^{'}} \right|^2 + \\
    & & 2 \sum_t \sum_{k,l \in L, k \neq l} 2^{k+l} \imath^{u_k(t)- u_l(t)}
    \cdot \\
    & & \hspace{0.8in} (\imath^{\kappa_k}-\imath^{\kappa_{k}^{'}})
    (\imath^{-\kappa_l}-\imath^{-\kappa_{l}^{'}}) \\
    & \geq & 2N \sum_{k \in L} 4^k \mid
    \imath^{\kappa_k}-\imath^{\kappa_{k}^{'}} \mid^2 - \\
    & & 2 \sqrt{N+1} \sum_{k,l \in L, k \neq l} 2^{k+l}
    (\imath^{\kappa_k}-\imath^{\kappa_{k}^{'}})
    (\imath^{\kappa_l}-\imath^{\kappa_{l}^{'}})
\eean For large $N$, this bound is minimized when \bean
    \kappa_k & = & \kappa_{k}^{'} \ \ \ \ \ \ 1 \leq k \leq (m-1) \\
    \kappa_0 & = & \kappa_0^{'} + 1 \ ,
\eean leading to \beqn
    d_{\min}^{\, 2} \ = \ 4 N .
\eeqn Upon normalization with the energy of the sequences, we obtain
\bean
    \overline{d}_{\min}^{\, 2} & \approx & \frac{4N}{\frac{2(M^2-1)}{3}} \ = \
    \frac{6N}{M^2-1} .
\eean

\section{Bounding the Correlation under Variable Rate Signalling for
Families $\{ {\cal CQ}_{M^2}\}$} \label{app:var_pf}

Let the sequences assigned to two users from Families ${\cal
CQ}_{M_1^2}$ and ${\cal CQ}_{M_2^2}$ be given by \beqan
    s(g, \kappa, t) & = & \sqrt{2 \imath} \,
    \sum_{k = 0}^{m_1 - 1} \, 2^k \, \imath^{u_k(t)} \imath^{\kappa_k}  \\
    s(g^{'}, \kappa^{'}, t) & = & \sqrt{2 \imath} \,
    \sum_{l = 0}^{m_2 - 1} \, 2^l \, \imath^{u_l^{'}(t)} \imath^{\kappa_l^{'}} .
\eeqan We assume, without loss of generality, that $m_1>m_2$. Then
\beqn
    {\theta_{s(g, \kappa), s(g^{'}, \kappa^{'})} (\tau) }
    \ = \  2 \sum_{k = 0}^{m_1 - 1} \sum_{l = 0}^{m_2 - 1}
    2^{k + l} \theta_{u_k, u_l^{'}}(\tau)\imath^{\kappa_k - \kappa_l^{'}}.
\eeqn Each $\theta_{u_k, u_l^{'}}(\tau)$ corresponds to correlation
of a pair of sequences from Family $\cal A$ and has magnitude
bounded by $(1 + \sqrt{N + 1})$.  Therefore, the magnitude of
$\theta_{s(g, \kappa), s(g^{'}, \kappa^{'})} (\tau)$ can be bounded
as \bean
    \left| \theta_{s(g, \kappa), s(g^{'}, \kappa^{'})} (\tau) \right|
    & \lesssim & 2(2^{m_1}-1)(2^{m_2}-1) \sqrt{N+1} \\
    & = & 2(M_1 - 1)(M_2 - 1) \sqrt{N+1}.
\eeqan The sequences $\{s(g, \kappa, t)\}$, $\{s(g^{'}, \kappa^{'},
t)\}$ have energy $\frac{2(M_1^2-1)}{3}$ and $\frac{2(M_2^2-1)}{3}$
respectively.   Upon normalization, we obtain the bound on
normalized correlation \bean
    \overline{\theta}_{s(g, \kappa), s(g^{'}, \kappa^{'})}(\tau)
    & \lesssim & \frac{2(M_1-1)(M_2-1)}{\frac{2}{3} \sqrt{(M_1^2-1)(M_2^2-1)}}\sqrt{N+1} \\
    & = & \frac{3\sqrt{(M_1-1)(M_2-1)}}{\sqrt{(M_1+1)(M_2+1)}}\sqrt{N+1} .
\eean The normalized non-trivial autocorrelation functions are
bounded respectively by \bean
    \overline{\theta}_{s(g, \kappa), s(g, \kappa)} (\tau)
    & \lesssim  & \frac{3(M_1-1)^2}{ (M_1^2-1)}\sqrt{N+1} \\
    & = &  \frac{3(M_1-1) }{ (M_1+1)} \sqrt{N+1},
\eean for sequences over QAM constellations of size $M_1^2$, and
\bean
    \overline{\theta}_{s(g^{'}, \kappa^{'}), s(g^{'}, \kappa^{'})} (\tau)
    & \lesssim &  \frac{3(M_2-1)}{(M_2+1)}\sqrt{N+1} ,
\eean for sequences over QAM constellations of size $M_2^2$.  It
follows that the normalized maximum correlation magnitude
$\overline{\theta}_{\max}$ experienced by the user with a smaller
constellation is given by \bean
    \overline{\theta}_{\max} & \lesssim &
    \frac{3\sqrt{(M_1-1)(M_2-1)}}{\sqrt{(M_1+1)(M_2+1)}}\sqrt{N+1}
\eean whereas the value of $\overline{\theta}_{\max}$ experienced by
the user with larger constellation size remains unchanged at \bean
    \overline{\theta}_{\max} & \lesssim &
    \frac{3(M_1-1)}{(M_1+1)}\sqrt{N+1}.
\eean

\section{Correlation of Sequences in Family ${\cal SQ}_{M^2}$}
\label{app:main_pq_corr_pf}

\noindent (Proof of Property (\ref{thm:main_pq_corr}) in
Theorem~\ref{thm:main_pq})

We establish properties for the basic sequences and leave the
straightforward extension of the results to the general case to the
reader.

For a fixed $M$, let $\{ s(g,  0, t) \}$ and $\{ s(g^{'}, 0, t) \}$
be two basic sequences from Family ${\cal SQ}_{M^2}$ where \beqa
    s(g, 0, t) & = & \sqrt{2 \imath} \, \sum_{k = 0}^{m - 1} \, 2^{m - k - 1} \,
        \imath^{u_k(t)} \label{eqref:s_g1}\\
    s(g^{'}, 0, t) & = & \sqrt{2 \imath} \, \sum_{l = 0}^{m - 1} \, 2^{m - l - 1} \,
        \imath^{u^{'}_l(t)} \label{eqref:s_g2}.
\eeqa with \beqa
    u_0(t) & = & T([1 + 2 g]\xi^t) \nonumber\\
    u_k(t) & = & T([1 + 2 (g + \delta_k)]\xi^{t + \tau_k}) \label{eqref:u_i}\\
    & & \hspace{1in} k = 1, 2, \ldots, m - 1 \nonumber \\
    u^{'}_0(t) & = & T([1 + 2 g^{'}]\xi^t) \nonumber\\
    u^{'}_k(t) & = & T([1 + 2 (g^{'} + \delta_k)]\xi^{t + \tau_k}) \label{eqref:u^{'}_i}\\
    & & \hspace{1in} k = 1, 2, \ldots, m - 1. \nonumber
\eeqa The correlation between the two sequences, $\{ s(g,  0, t) \}$
and $\{ s(g^{'},  0, t) \}$, is given by \beqa
    \lefteqn{\theta_{s(g, 0), \ s(g^{'}, 0)}(\tau)} \nonumber\\
    & = & \sum_{t = 0}^{N - 1} \, \left(
    \sqrt{2 \imath} \sum_{k = 0}^{m - 1} \, 2^{m - k - 1} \, \imath^{u_k(t + \tau)}
    \right) \cdot \nonumber\\
    & & \hspace{0.3in} \left( \overline{\sqrt{2 \imath}
    \sum_{l = 0}^{m - 1} \, 2^{m - l - 1} \, \imath^{u^{'}_l(t)} }
    \right) \nonumber\\
    & = & 2 \ \sum_{k, \, l = 0}^{m - 1} \, 2^{2m - k - l - 2} \,
    \left( \sum_{t = 0}^{N - 1} \imath^{u_k(t + \tau) - u^{'}_l(t)}
    \right) \nonumber\\
    & = & 2 \ \sum_{k, \, l = 0}^{m - 1} \, 2^{2m - k - l - 2}
    \theta_{u_k, \ u^{'}_l}(\tau). \label{eqref:cc}
\eeqa

The correlation properties of sequences in Family ${\cal SQ}_{M^2}$
are handled in two separate cases.

\subsubsection{Zero time shift $(\tau = 0)$}

Only the case $g \neq g'$ is of interest here.  Then, \beqn
    \theta_{s(g, 0), s(g^{'}, 0)}(0) \ = \
    2 \ \sum_{k, \, l = 0}^{m - 1} \, 2^{2m - k - l - 2}
    \theta_{u_k, \ u^{'}_l}(0).
\eeqn  For the case $k=l$ we have:  \bean
    \lefteqn{\theta_{u_k, \ u^{'}_k}(0)}\\
    & = & \sum_t
    \imath^{T([1 + 2(g_k + \delta_k)]\xi^{t + \tau_k}) -
    T([1 + 2(g^{'}_k + \delta_k)]\xi^{t + \tau_k})} \\
    & = & \sum_t \imath^{2T([g_k - g^{'}_k] \xi^{t + \tau_k})} \\
    & = & -1  \ \ (\text{see
    Appendix~\ref{app:closed_form_Fam_A}}).
\eean

For the case $k \neq l$, we have the bound \[ \left| \theta_{u_k,
\ u^{'}_l}(0) \right| \ \leq \ (1 + \sqrt{N + 1}) .\]  This leads
to  \bea
    \lefteqn{\left| \theta_{s(g), s(g^{'})}(0) \right|} \nonumber\\
    & \leq & 2(2^m - 1)^2 +
    2 \left( \frac{2}{3} 4^m - 2^{m+1} + \frac{4}{3} \right) \ \sqrt{N +
    1} \nonumber\\
    & \lesssim & 2 \left( \frac{2}{3} 4^m - 2^{m+1} + \frac{4}{3} \right) \
    \sqrt{N} . \label{eq:cc_zero}
\eea

\subsubsection{Non-zero time shift $(\tau \neq 0)$}

Here we need to consider, in addition, the case when $g=g'$.

We rewrite the expression for $\theta_{s(g, 0), \, s(g^{'},
0)}(\tau)$ from (\ref{eqref:cc}) to get \bea
    \lefteqn{\theta_{s(g, 0), \, s(g^{'}, 0)}(\tau) } \nonumber\\
    & = & 2 \ \left[ \sum_{k, l = 0, k \neq l}^{m - 1}
    \left( 2^{2(m - 1) - k - l}\theta_{u_k, \ u_l^{'}}(\tau) \right) +
    \right. \nonumber\\
    & & \hspace{0.35in} \left.
    \sum_{k = 0}^{m - 1} 2^{2(m - k - 1)} \theta_{u_k, \ u_k^{'}}(\tau) \right].
    \label{eqref:cc_non_zero}
\eea Using Lemma~\ref{lem:Fam_A} in
Appendix~\ref{app:closed_form_Fam_A}, we can rewrite the expressions
for $\theta_{u_k, \ u_k^{'}}(\tau)$ as: \beqn
    \theta_{u_k, \ u_k^{'}}(\tau) \ = \
    -1 + \Gamma(1) \imath^{-T(z_k)} \ , \ \ k = 0, 1, \ldots, m - 1
\eeqn where \bean
    z_0 & = &   g + \frac{g + g^{'}}{y} +
    \frac{1}{\sqrt{y}}  + 2 \mu(g, g^{'}, y) , \\
    z_k    & = & (g + \delta_k) +
    \frac{(g + g^{'})}{y} +
    \frac{1}{\sqrt{y}}  + 2 \mu(g + \delta_k, g^{'} + \delta_k, y)\\
    & = & z_0 + \delta_k + 2 \mu^{'}(g + \delta_k, g^{'} + \delta_k, y).
\eean The element $y$ is a function of the time shift $\tau$.

Substituting the expressions for $\theta_{u_k, \ u_k^{'}}(\tau)$ in
(\ref{eqref:cc_non_zero}), we get \beqan
    \lefteqn{\theta_{s(g, 0), \, s(g^{'}, 0)}(\tau) } \nonumber\\
    & = & 2 \ \left( \sum_{k, l = 0, k \neq l}^{m - 1}
    \left( 2^{2(m - 1) - k - l}\theta_{u_k, \ u_l^{'}}(\tau) \right) +
    \right. \nonumber\\
    & & \hspace*{0.4in} \sum_{k = 1}^{m - 1} 2^{2(m - k - 1)} \left( -1 + \right. \\
    & & \hspace*{0.4in} \left. \Gamma(1)
    \imath^{-T(z_0 + \delta_k + 2 \mu^{'}(g + \delta_k, g^{'} + \delta_k, y))} \right)  + \\
    & &  \hspace*{0.4in} \left. 2^{2m-2} \left( -1 + \Gamma(1) \imath^{-T(z_0)}
    \right)\right)\\
    & = & 2 \ \left( \sum_{k, l = 0, k \neq l}^{m - 1}
    \left( 2^{2(m - 1) - k - l}\theta_{u_k, \ u_l^{'}}(\tau) \right) +
    \right. \nonumber\\
    & & \hspace*{0.4in} \sum_{k = 1}^{m - 1} 2^{2(m - k - 1)}
    \left( -1 + \right. \\
    & & \hspace*{0.4in} \left. \Gamma(1) \imath^{-T(z_0)} \imath^{-T(\delta_k)}
    (-1)^{tr(\mu^{'}(g + \delta_k, g^{'} + \delta_k, y))} \right)  + \\
    & &  \hspace*{0.4in} \left.
    2^{2m-2}  \left( -1 + \Gamma(1) \imath^{-T(z_0)}
    \right)\right).
\eeqan We know that $\left|\theta_{u_k, \ u_l^{'}}(\tau)\right|$
can be bounded as $(1 + |\Gamma(1)|) \ = \ (1 + \sqrt{N+ 1})$.
Also, $tr(\delta_k) = 1$. It follows that  \bea
    \lefteqn{| \theta_{s(g, 0), \ s(g^{'}, 0)}(\tau) |} \nonumber\\
    & \leq & \left| 2(2^m - 1)^2 +
    2 \ \Gamma(1) \left( \sum_{k, l = 0, k \neq l}^{m - 1} 2^{2(m - 1) - k - l}
    + \right. \right. \nonumber\\
    & & \left. \left.
    \sum_{k = 0}^{m - 2} 2^{2k} (\imath) + 2^{2 m - 2} (1) \right) \right|
    \nonumber\\
    & = &  \left| 2(2^m - 1)^2 + 2 \, \Gamma(1) \left( 4^{m - 1} +
    \frac{2}{3} 4^m - 2^{m+1} + \right. \right. \nonumber\\
    & & \left. \left. \frac{4}{3} + \frac{4^{m - 1} - 1}{3} \imath \right)
    \right| \nonumber\\
    & \lesssim & \left(  \frac{61}{18} 16^m -
    \frac{44}{3} 8^m + \frac{230}{9} 4^m -
    \frac{64}{3} 2^m + \frac{68}{9}  \right)^\frac{1}{2} \, \sqrt{N}
    .\nonumber \\
    \label{eq:cc_nonzero_bd}
\eea

To determine $\theta_{max}$ (and hence $\overline{\theta}_{max}$ as
well), for Family ${\cal SQ}_{M^2}$, it turns out to be enough to
restrict attention to correlations amongst basic sequences. With the
aid of Property (\ref{thm:main_pq_energy}) dealing with the energy
of sequences of Family ${\cal SQ}_{M^2}$ and the correlation bounds
in \eqref{eq:cc_zero} and \eqref{eq:cc_nonzero_bd}, we arrive at the
following upper bounds on $\theta_{\max}$ and
$\overline{\theta}_{\max}$ for large $M,N$: \beqn
    \theta_{\max} \lesssim \frac{\sqrt{122}}{6} \, M^2 \, \sqrt{N}
\eeqn and \beqn
    \overline{\theta}_{\max} \ \lesssim \ \frac{\sqrt{122}}{4} \, \sqrt{N} .
\eeqn

\section{Correlation of Sequences in Family ${\cal IQ}_{16}$} \label{app:main_iq_16_corr_pf}

\noindent (Proof of Property (\ref{thm:main_iq_16_corr}) of
Theorem~\ref{thm:main_iq_16})

As in the case of Family ${\cal SQ}_{16}$, we define basic sequences
in Family ${\cal IQ}_{16}$ as sequences corresponding to assigning
$\kappa_0=\kappa_1=0$ in \eqref{eq:seq_fam_a}. We analyze the
correlation between two basic sequences from Family ${\cal IQ}_{16}$
and it is straightforward to extend the results to the case of
modulated sequences.

Let $\{ s(g_1,  0, t) \}$ and $\{ s(g_2, 0, t) \}$ be two basic
sequences belonging to Family ${\cal IQ}_{16}$, i.e.,  \beqa
    s(g_1, 0, t) & = & \left\{
    \begin{array}{c}
    \sqrt{2 \imath} \left( \imath^{u_1(t)} +
    2 \, \imath^{u_0(t)}  \right) \ , \ t \ \mbox{even} \\
    \sqrt{2 \imath}  \, \imath \left( \imath^{u_0(t)} -
    2 \, \imath^{u_1(t)}  \right) \ , \ t \ \mbox{odd}
    \end{array} \right.  \label{eqref:opt_s_g1} \\
    s(g_2, 0, t) & = & \left\{
    \begin{array}{c}
    \sqrt{2 \imath} \left( \imath^{v_1(t)} +
    2 \, \imath^{v_0(t)}  \right) \ , \ t \ \mbox{even} \\
    \sqrt{2 \imath}  \, \imath \left( \imath^{v_0(t)} -
    2 \, \imath^{v_1(t)}  \right) \ , \ t \ \mbox{odd}
    \end{array} \right. \label{eqref:opt_s_g2}
\eeqa where \beqan
    u_0(t) & = & T([1 + 2 \, g_1] \xi^t) \\
    u_1(t) & = & T([1 + 2 \, (g_1 + \delta_1)] \xi^{t + \tau_1}) \\
    v_0(t) & = & T([1 + 2 \, g_2] \xi^t) \ \ \mbox{and} \\
    v_1(t) & = & T([1 + 2 \, (g_2 + \delta_1)] \xi^{t + \tau_1}).
\eeqan

\blem \label{lem:cc_non_zero_opt} Let $\{ s(g_1,  0, t) \}$ and $\{
s(g_2,  0, t) \}$ be two sequences defined in \eqref{eqref:opt_s_g1}
and \eqref{eqref:opt_s_g2}. Then \beqn
    \theta_{s(g_1), s(g_2)}(\tau) \ \lesssim \ \sqrt{2} \, \sqrt{N} \ ,
    \ \ 0 \leq \tau < N.
\eeqn

\bpf The expression for the correlation between the two Q-PAM
sequences will take on one of two forms depending on if $\tau = 0
\ (\mbox{mod} \ 2)$ or if $\tau = 1 \ (\mbox{mod} \ 2)$.

Let us suppose that $\tau = 0 \ (\mbox{mod} \ 2)$.  In that case,
the correlation between the two sequences can be written as: \beqan
    \lefteqn{\theta_{s(g_1), s(g_2)}(\tau)}\\
    & = & \sum_{t = 0}^{N - 1} \, s(g_1, 0, t + \tau)
    \, \overline{s(g_2, 0, t)} \\
    & = & 2 \sum_{t \ \mbox{\tiny even}} \, \left( \imath^{u_1(t + \tau)} +
    2 \, \imath^{u_0(t + \tau)}  \right) \left( \imath^{-v_1(t)} +
    2 \, \imath^{-v_0(t)}  \right) + \\
    & &
    2 \sum_{t \ \mbox{\tiny odd}} \, \left( \imath^{u_0(t + \tau)} -
    2 \, \imath^{u_1(t + \tau)}  \right) \left( \imath^{-v_0(t)} -
    2 \, \imath^{-v_1(t)}  \right) \\
    & = & 2 \, \left( \ \theta_{u_1, \, v_1}(\tau) + 2 \, \theta_{u_0, \, v_1}(\tau) +
    2 \, \theta_{u_1, \, v_0}(\tau) + \right. \\
    & & \left. 4 \, \theta_{u_0, \, v_0}(\tau) \right) +
    2 \, \left( \ \theta_{u_0, \, v_0}(\tau) - 2 \, \theta_{u_0, \, v_1}(\tau) -
    \right. \\
    & & \left. 2 \, \theta_{u_1, \, v_0}(\tau) + 4 \, \theta_{u_1, \, v_1}(\tau) \right) \\
    & = & 10 \left( \theta_{u_0, \, v_0}(\tau) + \theta_{u_1, \, v_1}(\tau) \right)
\eeqan

Using the results from Appendix~\ref{app:main_pq_corr_pf}, we can
see that $\theta_{u_0, \, v_0}(\tau)$ and $\theta_{u_1, \,
v_1}(\tau)$ are at right angles to each other.  We can bound the
magnitude of $\theta_{s(g_1), s(g_2)}(\tau)$ in the above expression
as \beqa
    \left| \theta_{s(g_1),s(g_2)}(\tau) \right| & \leq & \left| 10 + 10 \, \Gamma(1) (1 +
    \imath) \right| \nonumber\\
    & \lesssim & 10 \, \sqrt{2} \, \sqrt{N/2} \nonumber\\
    & \lesssim & 10 \, \sqrt{N}. \label{eqref:bd1}
\eeqa

Now, if $\tau = 1 \ (\mbox{mod} \ 2)$, we get \beqa
    \lefteqn{\theta_{s(g_1), s(g_2)}(\tau)} \nonumber\\
    & = & \sum_{t = 0}^{N - 1} \, s(g_1, 0, t + \tau)
    \, \overline{s(g_2, 0, t)} \nonumber\\
    & = & 2 \, \imath \, \sum_{t \ \mbox{\tiny even}} \, \left( \imath^{u_0(t + \tau)} -
    2 \, \imath^{u_1(t + \tau)}  \right) \left( \imath^{-v_1(t)} +
    2 \, \imath^{-v_0(t)}  \right) - \nonumber\\
    & &
    2 \, \imath \, \sum_{t \ \mbox{\tiny odd}} \, \left( \imath^{u_1(t + \tau)} +
    2 \, \imath^{u_0(t + \tau)}  \right) \left( \imath^{-v_0(t)} -
    2 \, \imath^{-v_1(t)}  \right) \nonumber\\
    & = & 2 \, \imath \, \left( \ \theta_{u_0, \, v_1}(\tau) - 2 \, \theta_{u_1, \, v_1}(\tau) +
    2 \, \theta_{u_0, \, v_0}(\tau) - \right. \nonumber\\
    & & \left. 4 \, \theta_{u_1, \, v_0}(\tau) \right) -
    2 \, \imath \, \left( \ \theta_{u_1, \, v_0}(\tau) -
    2 \, \theta_{u_1, \, v_1}(\tau) \ + \right. \nonumber\\
    & & \left. 2 \, \theta_{u_0, \, v_0}(\tau) - 4 \, \theta_{u_0, \, v_1}(\tau) \right) \nonumber\\
    & = & 10 \, \imath \, \left( \theta_{u_0, \, v_1}(\tau) - \theta_{u_1, \, v_0}(\tau)
    \right).
    \label{eq:q1_16_odd_corr}
\eeqa The two correlations appearing in the above expression, viz.
$\theta_{u_0, \, v_1}(\tau)$ and $\theta_{u_1, \, v_0}(\tau)$ are
not aligned with respect to each other.  In the worst case, both of
them will contribute $(1 + \Gamma(1))$ to the final correlation
expression.  With that, we can bound the magnitude of
$\theta_{s(g_1), s(g_2)}(\tau)$ in the above expression as \beqa
    \left| \theta_{s(g_1),s(g_2)}(\tau) \right| & \leq &
    \left| 10 \, \Gamma(1) (1 + 1) \right| \nonumber\\
    & \lesssim & 20 \, \sqrt{N/2} \nonumber\\
    & \lesssim & 10 \sqrt{2} \, \sqrt{N}. \label{eqref:bd2}
\eeqa

With the bounds in \eqref{eqref:bd1} and \eqref{eqref:bd2}, and by
normalizing with the energy of the sequences, we get the statement
of the Lemma. \epf \elem

\section{Correlation of Sequences in Family ${\cal IP}_{8}$} \label{app:welch_bd_pf}

As with other sequence families and for the same reasons, we
analyze the correlation between two basic sequences in Family
${\cal IP}_{8}$ corresponding to the assignment
$\kappa_0=\kappa_1=0$ in \eqref{eq:eq_opt}.

Let $\{ s(g,  0, t) \}$ and $\{ s(g^{'}, 0, t) \}$ be two basic
sequences belonging to Family ${\cal IP}_{8}$, i.e.,  \beqa
    s(g, 0, t) & = & \left\{
    \begin{array}{c}
    \sqrt{2 \imath} \left( \imath^{u_1(t)} +
    2 \, \imath^{u_0(t)}  \right) \ , \ t \ \mbox{even} \\
    \sqrt{2 \imath}  \, \imath \left( \imath^{u_0(t)} -
    2 \, \imath^{u_1(t)}  \right) \ , \ t \ \mbox{odd}
    \end{array} \right.  \label{eqref:wel_s_g1} \\
    s(g^{'}, 0, t) & = & \left\{
    \begin{array}{c}
    \sqrt{2 \imath} \left( \imath^{u^{'}_1(t)} +
    2 \, \imath^{u^{'}_0(t)}  \right) \ , \ t \ \mbox{even} \\
    \sqrt{2 \imath} \, \imath \left( \imath^{u^{'}_0(t)} -
    2 \, \imath^{u^{'}_1(t)}  \right) \ , \ t \ \mbox{odd}
    \end{array} \right. \label{eqref:wel_s_g2}
\eeqa where \beqan
    u_0(t) & = & T([1 + 2 \, g] \xi^t) \\
    u_1(t) & = & T([1 + 2 \, (g + \delta_1)] \xi^t) \\
    u^{'}_0(t) & = & T([1 + 2 \, g^{'}] \xi^t) \ \ \mbox{and} \\
    u^{'}_1(t) & = & T([1 + 2 \, (g^{'} + \delta_1)] \xi^t).
\eeqan

The expression for the correlation between the two Q-PAM sequences
will take on one of two forms depending on if $\tau = 0 \
(\mbox{mod} \ 2)$ or if $\tau = 1 \ (\mbox{mod} \ 2)$.

Let us suppose that $\tau = 0 \ (\mbox{mod} \ 2)$.  In that case,
the correlation between the two sequences can be written as (see
Appendix~\ref{app:main_iq_16_corr_pf} for details): \beqan
    \theta_{s(g), s(g^{'})}(\tau) \ = \
    10 \left( \theta_{u_0, \, u^{'}_0}(\tau) + \theta_{u_1, \, u^{'}_1}(\tau)
    \right).
\eeqan

Using the results from Appendix~\ref{app:main_pq_corr_pf}, we can
see that $\theta_{u_0, \, u^{'}_0}(\tau)$ and $\theta_{u_1, \,
u^{'}_1}(\tau)$ are at right angles to each other.  We can bound the
magnitude of $\theta_{s(g), s(g^{'})}(\tau)$ in the above expression
as \beqa
    \left| \theta_{s(g),s(g^{'})}(\tau) \right| & \leq & \left| 10 + 10 \, \Gamma(1) (1 +
    \imath) \right| \nonumber\\
    & \lesssim & 10 \, \sqrt{2} \, \sqrt{N/2} \nonumber\\
    & \lesssim & 10 \, \sqrt{N}. \label{eqref:wel_bd1}
\eeqa

Now, if $\tau = 1 \ (\mbox{mod} \ 2)$, we get (see
Appendix~\ref{app:main_iq_16_corr_pf} for details) \beqa
    \theta_{s(g), s(g^{'})}(\tau) \ = \
    10 \, \imath \, \left( \theta_{u_0, \, u^{'}_1}(\tau) - \theta_{u_1, \, u^{'}_0}(\tau)
    \right).
    \label{eq:wel_16_odd_corr}
\eeqa From the results in Appendix~\ref{app:closed_form_Fam_A}, the
two correlations appearing in the above expression, viz.
$\theta_{u_0, \, u^{'}_1}(\tau)$ and $\theta_{u_1, \,
u^{'}_0}(\tau)$ can be rewritten as \bean
    \theta_{u_0, \, u^{'}_1}(\tau) & = & -1 + \Gamma(1)
    \imath^{-T(z_0)} \\
    \theta_{u_1, \, u^{'}_0}(\tau) & = & -1 + \Gamma(1)
    \imath^{-T(z_1)} \ ,
\eean where \bean
    z_0 & = & g + \frac{g + g^{'} + \delta_1}{y} + \frac{1}{\sqrt{y}} +
    2 \mu(g, g^{'} + \delta_1, y) \\
    z_1 & = & g + \delta_1 + \frac{g + \delta_1 + g^{'}}{y} + \frac{1}{\sqrt{y}} +
    2 \mu(g + \delta_1, g^{'}, y) \\
    & = & z_0 + \delta_1 + 2 \mu^{'}(g + \delta_1, g^{'}, y).
\eean Since $tr(\delta_1) = 1$, we can see that the two correlation
terms appearing in \eqref{eq:wel_16_odd_corr} are at right angles
and we can bound the magnitude of $\theta_{s(g), s(g^{'})}(\tau)$ as
\beqa
    \left| \theta_{s(g),s(g^{'})}(\tau) \right| & \leq & \left| 10 + 10 \, \Gamma(1) (1 +
    \imath) \right| \nonumber\\
    & \lesssim & 10 \, \sqrt{2} \, \sqrt{N/2} \nonumber\\
    & \lesssim & 10 \, \sqrt{N}. \label{eqref:wel_bd2}
\eeqa

With the bounds in \eqref{eqref:wel_bd1} and \eqref{eqref:wel_bd2},
and by normalizing with the energy of the sequences, we can bound
$\theta_{\max}$ and $\overline{\theta}_{\max}$ for Family ${\cal
IP}_{8}$ as: \bean
    \theta_{\max} & \lesssim & 10 \, \sqrt{N} \\
    \overline{\theta}_{\max} & \lesssim & \sqrt{N}.
\eean

\bibliographystyle{IEEEbib}

\begin{biographynophoto}{M. Anand}
received the B.E. degree in Electronics and Communication
Engineering from the Visvesvaraya Technological University, Belgaum,
in 2003 and the M.Sc.(Engg.) degree from the Indian Institute of
Science, Bangalore, in 2007.

He is currently a Ph.D. student at the Coordinated Science
Laboratory, University of Illinois, Urbana, IL. His research
interests include multi-user information theory and code design for
wireless communications.
\end{biographynophoto}

\begin{biographynophoto}{P. Vijay Kumar}
(S'80-M'82-SM'01-F'02) received the B.Tech. and M.Tech. degrees from
the Indian Institutes of Technology (Kharagpur and Kanpur), and the
Ph.D. Degree from the University of Southern California (USC), Los
Angeles, in 1983, all in electrical engineering.  Since 1983 he has
been on the faculty of the EE-Systems Department of USC.  He is
presently on leave of absence from USC at the Indian Institute of
Science, Bangalore.  His research interests include space-time codes
for cooperative communication networks,
 low-correlation sequences for wireless and optical CDMA and
sensor networks. A low-correlation sequence family co-designed by
him is now part of the 3G-WCDMA standard.
 He was an Associate Editor for Coding Theory for the IEEE
Transactions on Information Theory, 1993-1996. In 1994, he received
the USC School-of-Engineering Senior Research Award for
contributions to coding theory.  He is a co-recipient of the IEEE
Information Theory Society 1995 Prize Paper Award.
\end{biographynophoto}
\end{document}